\documentclass[12pt,a4paper]{article}
\usepackage{ifpdf}
\usepackage{times}
\usepackage[scale={0.8,0.9},centering,includeheadfoot]{geometry}
\usepackage{enumitem}
\usepackage{subcaption}
\usepackage{multicol}

\usepackage{pgf}

\usepackage{amsmath,amssymb,xspace,amsthm,bm}

\usepackage[colon,longnamesfirst]{natbib}

\usepackage{dcolumn}\newcolumntype{d}[1]{D{.}{.}{#1}}

\def\qedbox{\ifvmode\else\unskip\fi~\penalty10000
    \hfill{\large$\blacksquare$}}

\def\miscinfo#1#2{{\footnotesize\indent\textsc{#1: }\ignorespaces #2}}   

\begin{document}
\bibliographystyle{apalike}

\title{
A unified view on Bayesian varying coefficient models}
\author{
    Maria Franco-Villoria \\
    Department of Economics and Statistics\\
    University of Torino\\
    \and
    Massimo Ventrucci \\
	Department of Statistical Sciences\\    
    University of Bologna\\
    \\
    \and
    H{\aa}vard Rue\\
    CEMSE Division, King Abdullah University of Science and Technology, \\
    Thuwal, Saudi Arabia
    }

\date{\today}

\maketitle

\begin{abstract}
Varying coefficient models are useful in applications where the effect of the covariate might depend on some other covariate such as time or location. Various applications of these models often give rise to case-specific prior distributions for the parameter(s) describing how much the coefficients vary. In this work, we introduce a unified view of varying coefficients models, arguing for a way of specifying these prior distributions that are coherent across various applications, avoid overfitting and have a coherent interpretation. We do this by considering varying coefficients models as a flexible extension of the natural simpler model and capitalising on the recently proposed framework of penalized complexity (PC) priors. We illustrate our approach in two spatial examples where varying coefficient models are relevant.
\end{abstract}

\miscinfo{Keywords}{INLA; overfitting; penalized complexity prior; varying coefficient models}\\
\miscinfo{Address for Correspondence}{Maria Franco-Villoria, Department of Economics and Statistics Cognetti de Martiis, University of Torino. Email: maria.francovilloria@unito.it}

\section{Introduction}
\label{sec:intro}

Varying coefficient models (VCMs, \cite{Hastie:1993}) can be seen as a general class of models that encompasses a large number of statistical models as special cases: the generalized linear model, generalized additive models, dynamic generalized linear models or even the more recent functional linear models. They can also be seen as a particular case of structured additive regression (STAR) models \citep{Fahrmeir04penalizedstructured}. 

VCMs arise in a vast range of applications, ranging from economics to epidemiology \citep{Gelfand:2003, Hoover:1998, Ferguson:2007, Finley:2011, Mu:2018, Fan:1999, Cai:2003, Tian:2005}. In practice, varying coefficient models are useful in presence of an \emph{effect modifier}, a variable that ``changes'' the effect of a covariate of interest on the response. For the sake of a general notation that includes all cases discussed in this paper, consider the triplet $(y_t,x_t,z_t)$, $t=1,...,n$, observed on $n$ observational units, with $z$ being the variable modifying the relationship between the covariate $x$ and the response $y$. 
Following \cite{Hastie:1993} who introduce VCMs as an extension of generalized linear models \citep{Nelder:1972}, we assume $y$ belonging to the exponential family and model the effect of covariate $x$ in the scale of the \emph{linear predictor} $\eta = g(\mu)$, which is linked to the mean response $\mu$ via the \emph{link function} $g$. The linear predictor of a generalized VCM is
\[
\eta_t = \alpha +  \beta(z_t) x_{t} \ \ \ \ \ \ \ t=1,...,n,
\]
where $\beta(z_t)$, $t=1,...,n$, is the varying regression coefficient (VC), that can be regarded as a stochastic process on the effect modifier domain. For ease of notation we will use $\beta_t$ to denote $\beta(z_t)$. 

Depending on the nature of the effect modifier, that can be either a continuous variable (e.g. temperature) or a time/space index (e.g. day or municipality), we can envision several models for the varying coefficient $\boldsymbol \beta=(\beta_1,\ldots,\beta_n)^T$. For instance, we can assume exchangeability over $t=1,\ldots,n$ with $cor(\beta_i,\beta_j)=\xi$ for $i\neq j$ if there is no natural ordering among the values of $z$. If the effect modifier is time, $\beta_t$ might be a $1^{st}$ order autoregressive (AR1) and $\xi$ the lag-one correlation, or a spline if we want to ensure smoothness. The coefficients may also vary in space in a continuous or discrete way, in which case a Gaussian random field with a certain covariance function or a conditionally autoregressive (CAR) model can be assumed, respectively. 
These models have been treated separately in the literature, along with priors specifically chosen for each different model (see e.g. \cite{Biller:2001, Gamerman:2003, Gelfand:2003}). Further, the choice of the model describing the behaviour of the varying coefficient and the prior assigned to its parameter(s) (that controls the flexibility of the VC), are usually made at two different stages.

In this work we argue that 1) regardless of the model assumed for $\boldsymbol \beta$, all models can be seen in a unified way as a flexible extension of a simpler model where the varying coefficient is instead constant, and 2) the prior can be specified coherently with this model conception so that the issue of model and prior choice is tackled jointly. Following these two points, we propose a unified view on varying coefficient models where the prior is built under the recently proposed Penalized Complexity (PC) Prior framework \citep{pcprior}.

The flexibility that VCMs offer can be desirable in certain applications and much work has been devoted to the development of flexible models.
In our view, the VCM arises naturally from a simpler model; 
i.e. we can consider increasing the flexibility of the simple linear regression model $\eta_t = \alpha +  \beta x_{t}$, $t=1,\ldots,n$ by allowing the coefficient $\beta$ to vary over $t$. Common choices of the prior might lead to overfitting, i.e. might \emph{push} the model away from the simpler model even when a more flexible one is not appropriate \citep{FruhwirthSchnatter-2010, FruhwirthSchnatter-2011, pcprior}. The importance of using priors for VCMs that avoid overfitting is now beginning to be acknowledged in the literature by some authors such as \cite{Bitto:2018} and \cite{Kowal:2018}, who consider shrinkage priors for the variance parameter in a time-varying setting, the latter allowing to adapt the shrinkage locally.
We propose to use the more general PC prior approach to define priors that guarantee shrinkage to a simpler model for any kind of hyperparameter(s), including variance parameters. By treating varying coefficient models in a unified way we can use a single approach for doing so. 
The Penalized Complexity (PC) Prior framework considers a model component as a flexible extension of a simpler version of the model component, referred to as the base model. PC priors are defined on the scale of the distance from the base model and then transferred to the scale of the original parameter by a standard change of variable transformation. This strategy can be applied independently of the model choice  for $\boldsymbol \beta$ describing the VC in a unique way, as the base model can always be easily identified in terms of a value for $\xi$. In this sense, we propose a coherent framework for building varying coefficient models.

The plan of the paper is as follows. Section~\ref{sec:vcm} presents varying coefficient models in a unified way, while the general framework to construct PC priors is briefly reviewed in Section~\ref{sec:pcprior}.
In Section~\ref{sec:pc:rho}, several PC priors for $\xi$ are derived under different model choices for $\boldsymbol \beta$, focusing first on the unstructured case (Section~\ref{sec:pc:unstructured}), where the realizations of the VC are assumed to be exchangeable. Structured cases, such as time and space are presented in Sections~\ref{sec:pc:time} and \ref{sec:pc:space}. Section~\ref{sec:properties} discusses the properties of PC priors and how these compare to other priors. Examples are illustrated in Section~\ref{sec:examples}. The paper closes with a discussion in Section~\ref{sec:discussion}.

\section{A unified view on varying coefficient models}
\label{sec:vcm}

Let us now specify a Bayesian generalized VCM, seeing it as a flexible extension of the simple generalized linear model $\eta_{t} = \alpha + \beta_0 x_t$, which will be denoted as \emph{base model}; 
this can be thought of as the fit obtained if data do not show evidence for a varying coefficient but for a constant regression coefficient instead. Without loss of generality, we can assign the prior $\beta_0\sim \mathcal{N}(0,1)$ to the base model:
\begin{equation}
\begin{array}{l}
\eta_t = \alpha +  \beta_0 x_{t} \ \ \ \ \ \ \ t=1,...,n, \\
\beta_0 \sim N(0,1).
\end{array}
\label{eq:vcm-base}
\end{equation}
If we believe that the covariate effect is not constant in $z$, we can allow for deviation from $\beta_0$ in the form of a varying coefficient model,
\begin{equation}
\begin{array}{l}
\eta_t = \alpha +  \left(\beta_0 + \beta_t \right) x_{t} \ \ \ \ \ \ \ t=1,...,n,\\
\bm \beta|\xi \sim \pi(\bm\beta| \xi),
\end{array}
\label{eq:vcm-beta}
\end{equation}
where $\bm \beta=(\beta_1,\ldots,\beta_n)^T$ is a vector of random effects defining a stochastic process over $z$, denoted as $\pi(\bm \beta|\xi)$ with $\xi$ the associated hyperparameter(s). 

In what follows we will assume the linear predictor $\eta_t = \alpha +  \left(\beta_0 + \beta_t \right) x_{t}$ in Eq.~(\ref{eq:vcm-beta}) and consider different Gaussian models for $\pi(\bm \beta| \xi)$. We will focus on the VC models mostly used in applications, each of them representing a specific \emph{extension} of the base model in Eq.~(\ref{eq:vcm-base}), allowing us to view all the various cases in a unified manner.

\subsection{The unstructured case}
\label{sec:unstructured}
The simplest correlation structure for random effects is to assume that they are exchangeable; this is commonly used to account for dependence among repeated measures in longitudinal models \citep{Laird:1982}.
If $\bm \beta = (\beta_1, ..., \beta_n)^{\textsf{T}}$ are exchangeable over $t=1,\ldots,n$, then $\bm \beta \sim \mathcal{N}(0,\tau^{-1}\bm R(\tilde{\rho}))$, where the correlation matrix is
\begin{equation}
\bm R(\tilde{\rho}) = \left[
\begin{array}{cccccc}
  1 & \tilde{\rho} & \ldots  &       & \tilde{\rho}   \\
  \tilde{\rho} & 1  & \tilde{\rho} & \ldots    & \tilde{\rho}   \\
   \cdot &    &   \cdot&   &  \cdot   \\
   \cdot &    &  &   \cdot&  \cdot \\
    \tilde{\rho} &  \tilde{\rho}    & \ldots &  \tilde{\rho} & 1 \\
\end{array}
\right]
\label{eq:R-unstr}
\end{equation}
and $\tau$ is a precision parameter. For $\bm R(\tilde{\rho})$ to be positive definite, $-1/(n-1) < \tilde{\rho} < 1$ \citep{pcprior}. In the following, we consider $0 \leq \tilde{\rho}< 1$.

In this case, Model~(\ref{eq:vcm-beta}) can be reparametrized as $\eta_t=\alpha+\beta_tx_t$, $t=1,\ldots,n$ with
\begin{equation}\label{eq:unstr}
\bm \beta \sim N(0,\bm R(\rho)),
\end{equation}
assuming unit marginal variance with no loss of generality (in practical applications $\tau$ can either be fixed to a single value if known or it could be considered as a parameter on which we impose a prior distribution).
A sensible base model is $\rho=1$, corresponding to $\beta_t=\beta \ \ \forall t$.

\subsection{The structured case: temporal variation}
\label{sec:time} 
In many real life applications the values of the effect modifier follow a natural ordering, e.g. time, so that it is not realistic to assume exchangeability of $\beta_t$. Instead, autoregressive (AR) models from time series analysis can be adopted \citep{sorbye-2016}. An alternative is to consider the varying coefficient as a smooth function.
A popular model in the context of smoothing with splines is the $2^{nd}$ order random walk (RW2),  that can be seen as a discrete representation of a continous (integrated) Wiener process that retains the Markov property and is computationally efficient \citep{lindgren-2008}. It is also used in P-splines \citep{marx-vcm} where a RW2 is assigned to the coefficients of local B-spline basis functions. 
In the following we consider three cases: the $1^{st}$ order autoregressive (AR1) and the $1^{st}$ and $2^{nd}$ order random walk (RW1, RW2). In all three cases, we always assume the linear predictor reported in Eq.~(\ref{eq:vcm-beta}), but consider different models for $\beta_t$.

\subsubsection{The autoregressive model of first order}
\label{sec:AR1} 
The most common model for dependence on time is the autoregressive process of first order (AR1), the discrete-time analogue of the Ornstein-Uhlenbeck process, characterized by a correlation function with exponential decay rate.
A $1^{st}$ order autoregressive prior on the varying coefficient is $\beta_t = \tilde{\rho} \beta_{t-1} + w_t$, where $|\tilde{\rho}|<1$ represents the lag-one correlation, $w_t \sim \mathcal{N}(0,\tau^{-1}(1-\tilde{\rho}^2)), t=2,...,n$, and $\beta_1 \sim \mathcal{N}(0,\tau^{-1})$. 
The varying coefficient has a joint distribution given by $\bm \beta \sim \mathcal{N}\left(0,\tau^{-1}\bm R(\tilde{\rho})\right)$
with $\bm R(\tilde{\rho})_{ij}=(\tilde{\rho}^{|i-j|})$ and $\tau$ a precision parameter. 
Similarly to Section~\ref{sec:unstructured}, we can reparametrize Model~(\ref{eq:vcm-beta}) as $\eta_t=\alpha+\beta_tx_t$, $t=1,\ldots,n$, so that 
\begin{equation}\label{eq:R-AR1}
\bm \beta \sim \mathcal{N}\left(0,\bm R(\rho)\right)
\end{equation}
and $\beta_1 \sim N(0,1)$. In this case the base model is $\rho=1$, i.e. no change in time.

\subsubsection{Random walk model of order one and two}
\label{sec:rw}

We can consider the varying coefficient $\bm \beta$ in Eq.~(\ref{eq:vcm-beta}) as a smooth stochastic process on the effect modifier scale. The equivalence between smoothing splines and Gaussian processes was shown in \cite{Wahba:1970}. In a Bayesian framework, smoothing models are obtained using a random walk on the varying coefficients. A random walk is an intrinsic Gaussian Markov Random Field (IGMRF, \cite{rue-2005} ch. 3), i.e. a process with the multivariate Gaussian density 
\begin{equation}
\pi(\bm \beta|\tau) = (2\pi)^{-\texttt{rank}(\bm K)/2} (|\tau \bm K|^{*})^{1/2} \exp\left\{-\frac{\tau}{2}\bm \beta^{\textsf{T}} \bm K \bm \beta\right\}
\label{eq:igmrf}
\end{equation} 
where the structure matrix $\bm K$ is sparse and rank deficient ($\text{rank}(\bm K)=n-r$), $\tau$ is a scalar precision parameter and $|\tau \bm K|^{*}$ is the generalized determinant. 

The structure matrix encodes the conditional dependencies among the coefficients $\bm \beta$. To avoid scaling issues inherent in RW models, such as dependence on the graph, \cite{sorbye-2013} propose to scale the matrix $\bm K$ by a factor equal to the geometric mean of the diagonal elements of the generalized inverse of $\bm K$, so that the marginal variance (subject to appropriate sum to zero constraints) is equal to $1$. The rank deficiency of the structure matrix also identifies the order $r$ of the IGMRF. Model~(\ref{eq:igmrf}) describes deviation from a polynomial model of degree $r - 1$: e.g. a constant for RW1 ($r=1$) and a linear trend for RW2 ($r=2$). This means we need to impose a sum to zero constraint on $\bm \beta$ to avoid confounding with $\beta_0$ in Eq.~(\ref{eq:vcm-beta}), with the difference that using a RW2 will result in a smoother fit than if a RW1 is used. Without loss of generality, we assume equally spaced locations. The case of irregularly spaced locations differs only in the structure matrix $\bm K$ and the constraint, that has to be modified with the inclusion of appropriate integration weights \citep{lindgren-2008}.

The precision parameter $\tau$ regulates the amount of shrinkage towards the base model, that corresponds to $\tau=\infty$.

\subsection{The structured case: spatial variation}
\label{sec:space}

Spatially structured models include the cases of continuous or discrete spatial variation. In the former case, the effect modifier is the pair of (scaled) latitude and longitude coordinates, $\bm z_t=\{\text{lat}_t,\text{lon}_t\}$ and $\beta_t$ can be assumed as a realization from a spatial process. The class of Gaussian Random Field (GRF) models equipped with a \emph{Mat\'{e}rn} covariance is the most popular model \citep{Stein:1999}.
For areal data, the spatial units are identified by a one-dimensional region index, with no unique ordering among the regions. Neighbouring regions are assumed to be correlated, and the neighbourhood structure can be coded into a structure matrix. To model $\beta_t$, the standard approach is to use conditionally autoregressive (CAR) models proposed by \cite{besag}; see \cite{Waller:2007, Staubach:2002} for applications.

\subsubsection{Areal spatial variation}
\label{sec:areal}

Models for areal data have been widely discussed in the literature and are useful, for example, in epidemiological studies \citep{Banerjee:2015}, where data are not available at individual level but only at some aggregated level such as municipality or zip code (see Figure~\ref{fig:ex2-fit} for an example).
 
Assume the linear predictor in (\ref{eq:vcm-beta}) where $t=1,\ldots,n$ indicates each of the non overlapping regions in a lattice. Areas $i$ and $j$ are considered as neighbours, denoted as $i\sim j$, if they share a common border. The spatially varying coefficient $\bm \beta=(\beta_1, ..., \beta_n)^{\textsf{T}}$ follows an Intrinsic Conditional Autoregressive (ICAR) model \citep{besag}:
\[
	\beta_t|\bm \beta_{-t},\tau\sim \mathcal{N}\left(\frac{1}{n_{t}}\sum_{j:t\sim j}\beta_j,{(n_{t}\tau)}^{-1}\right)
\] 
with $n_{t}$ the number of neighbours of region $t$ and $\tau$ a precision parameter. The joint distribution for $\bm \beta$ is
\begin{equation}
\pi(\bm \beta|\tau) = (2\pi)^{-(n-1)/2} (|\tau \bm K|^{*})^{1/2} \exp\left\{-\frac{\tau}{2}\bm \beta^{\textsf{T}} \bm K \bm \beta\right\}
\label{eq:icar}
\end{equation}
where the structure matrix $\bm K$ is singular with null space $\bm 1$ and entries:
\[
	K_{i,j}=\left\{
	\begin{array}{ll}
	n_i & i=j\\
	-1 & i\sim j\\
	0 & \text{otherwise.}
	\end{array}
	\right.
\]
The base model, corresponding to no variation over area, is $\tau=\infty$.

\subsubsection{Continuous spatial variation}
\label{sec:geostats}
In this case, $t=(\text{lat}_t,\text{lon}_t)$, properly scaled, represents location within a spatial region $D \subseteq \mathbb{R}^2$ and the spatially varying coefficient can be seen as a realization of a Gaussian random field (GRF) with a \emph{Mat\'{e}rn} covariance function characterized by the marginal variance $\tau^{-1}$ and range parameter $\phi$. 
These two parameters cannot be estimated consistently under infill asymptotics \citep{Warnes:1987, Zhang:2004}, but only a function of those such as the product or the ratio, depending on the smoothness of the GRF. 

Assuming the linear predictor in (\ref{eq:vcm-beta}) the spatially varying coefficient 
\begin{equation}\label{eq:R-matern}
\bm \beta \sim \mathcal{N}\left(\bm 0,\tau^{-1}\bm R(\phi)\right)
\end{equation}
with $\bm R(\phi)_{ij}=(C({||i-j||)})$, $C(\cdot)$ is a \emph{Mat\'{e}rn} correlation function with fixed smoothness $\nu$:
\[
	C(h)=\frac{2^{1-\nu}}{\Gamma(\nu)}\left(\frac{\sqrt{8\nu}h}{\phi}\right)^{\nu}K_{\nu}\left(\frac{\sqrt{8\nu}h}{\phi}\right),
\]
$K_{\nu}$ is the modified Bessel function of second kind and order $\nu$ and $h$ represents the distance between any pair of locations. The base model in this case corresponds to $\tau=\infty$, $\phi=\infty$.

\section{Review of Penalized Complexity (PC) Priors} 
\label{sec:pcprior}
The prior for all the hyperparameters in Section~\ref{sec:vcm} can be built in a coherent way regardless of the assumed model for $\bm \beta$ using penalized complexity priors. 
In this section we summarize the four main principles underpinning the construction of PC priors, namely: support to Occam's razor (parsimony), penalisation of model complexity, constant rate penalisation and user-defined scaling. For a more detailed presentation of these principles the reader is referred to \cite{pcprior}. 

Let $f_1$ denote the density of a model component $\bm w$ where $\xi$ is the parameter for which we need to specify a prior. The base model, corresponds to a fixed value of the parameter $\xi=\xi_0$ and is characterized by the density $f_0$. 
\begin{enumerate}
	\item The prior for $\xi$  should give proper shrinkage to $\xi_0$ and decay with increasing complexity of $f_1$ in support of Occam's razor, ensuring parsimony; i.e. the simplest model is favoured unless there is evidence for a more flexible one. 

	\item The increased complexity of $f_1$ with respect to $f_0$ is measured using the Kullback-Leibler divergence \citep[KLD, ][]{kld-1951},
\[
	\text{KLD}(f_1||f_0)=\int f_1(w)\log\left(\frac{f_1(w)}{f_0(w)}\right)dw,
\]
For ease of interpretation, the KLD is transformed to a unidirectional distance measure
\begin{equation}\label{eq:KLD}
	d(\xi)=d(f_1||f_0)=\sqrt{2\text{KLD}(f_1||f_0)}
\end{equation}
that can be interpreted as the distance from the flexible model $f_1$ to the base model $f_0$. 

	\item The PC prior is defined as an exponential distribution on the distance, 
\begin{equation}
\pi(d(\xi)) = \lambda \exp(-\lambda d(\xi)),
\label{eq:pc}
\end{equation}	
	with rate $\lambda>0$. 
	The PC prior for $\xi$ follows by a change of variable transformation.
	\item The user must select $\lambda$ based on his prior knowledge on the parameter of interest (or an interpretable transformation of it $Q(\xi)$). This knowledge can be expressed in terms of a probability statement, e.g. $\mathbb{P}(Q(\xi)>U)=a$, where $U$ is an upper bound for $Q(\xi)$ and $a$ is a (generally small) probability.
\end{enumerate}

\section{PC priors for varying coefficient models}
\label{sec:pc:rho}

The PC prior framework offers a unified approach for constructing priors for all the various models considered in Section~\ref{sec:vcm} while guaranteeing proper shrinkage to the base model (\ref{eq:vcm-base}). Within this framework, we can always build the prior for the corresponding flexibility parameter(s) as an exponential distribution on the distance from the base model, then transform it back to the original scale. 
Here we present the main results about the derivation of PC priors for the flexibility parameters of the models considered in Section~\ref{sec:vcm}, while technical details can be found in the appendix.

\subsection{The unstructured case}
\label{sec:pc:unstructured}
As described in Section~\ref{sec:unstructured}, the base model for Model~(\ref{eq:unstr}) is $\rho=1$. The PC prior for $\rho$:

\begin{equation}
\label{eq:pc-rho-exch}
	\pi(\rho)=\frac{\theta\exp(-\theta\sqrt{1-\rho})}{2\sqrt{1-\rho}(1-\exp (-\theta))}, \quad 0\leq \rho<1, \quad \theta>0.
\end{equation}
 
The prior is scaled in terms of $\theta$ based on the prior belief that the user has about the parameter $\rho$ in the form of $(U,a)$ such that $\mathbb{P}(\rho>U)=a$. The corresponding value for $\theta$ is given by the solution of the equation
\[
	\frac{1-\exp(-\theta\sqrt{1-U})}{1-\exp(-\theta)}=a
\]
that has to be solved numerically, provided that $a>\sqrt{1-U}$. The PC prior in (\ref{eq:pc-rho-exch}) is illustrated in Figure~\ref{fig:pc-rho-exch} left panel.

\subsection{The structured case: temporal variation}
\label{sec:pc:time}
\subsubsection{The autoregressive model of first order}
For Model~(\ref{eq:R-AR1}), \cite{sorbye-2016} derive the PC prior with base model at $\rho=1$ as
\begin{equation}\label{eq:pc-rho-ar1}
\pi(\rho) = \frac{\theta \exp(-\theta \sqrt{1-\rho})}{2\sqrt{1-\rho}(1-\exp(-\sqrt{2}\theta)) }, \ \ \ \ \ \ \ \ \ \ |\rho| <1 , \quad \theta>0. 
\end{equation}

\begin{figure}
\centerline{
\includegraphics[width=.5\textwidth]{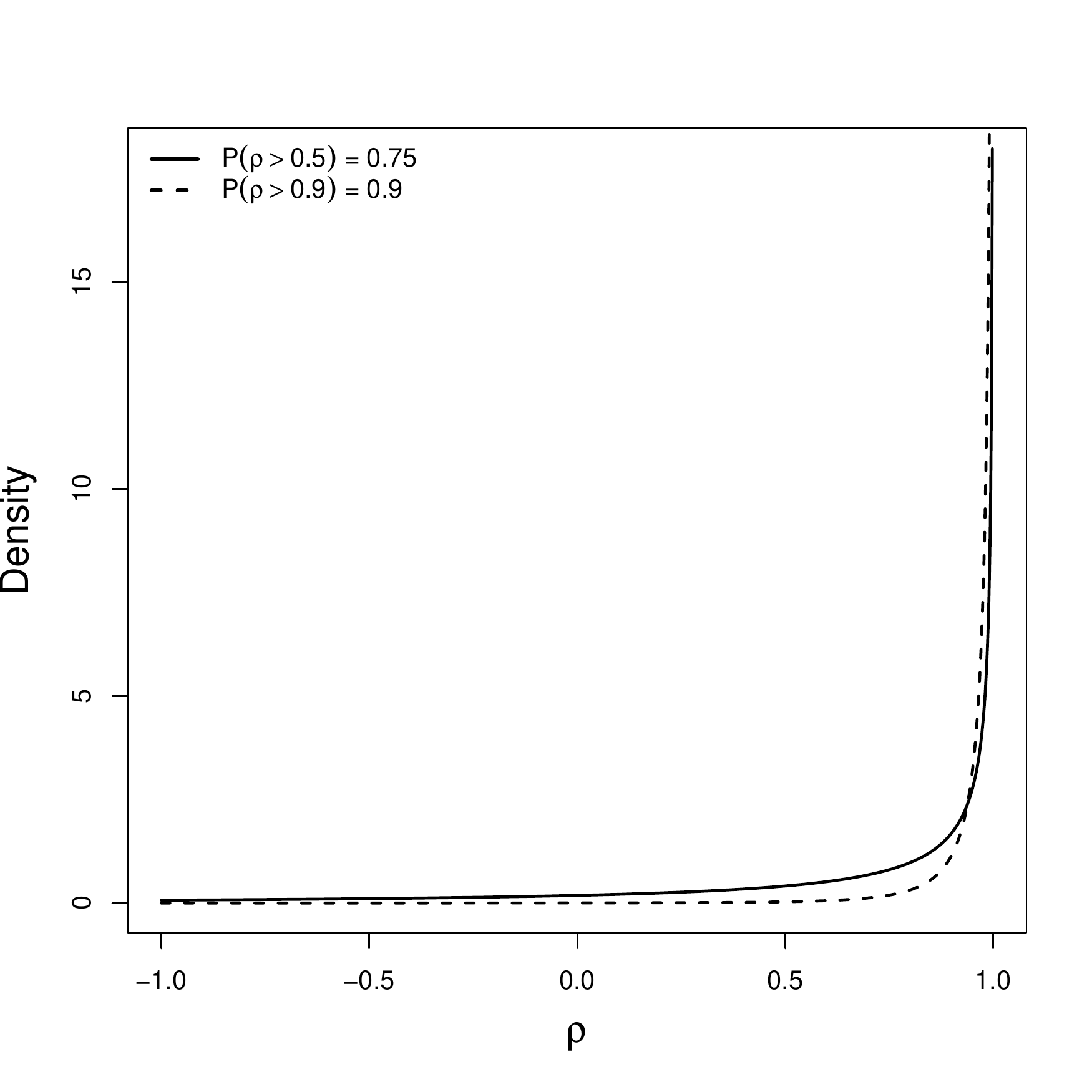}
\includegraphics[width=.5\textwidth]{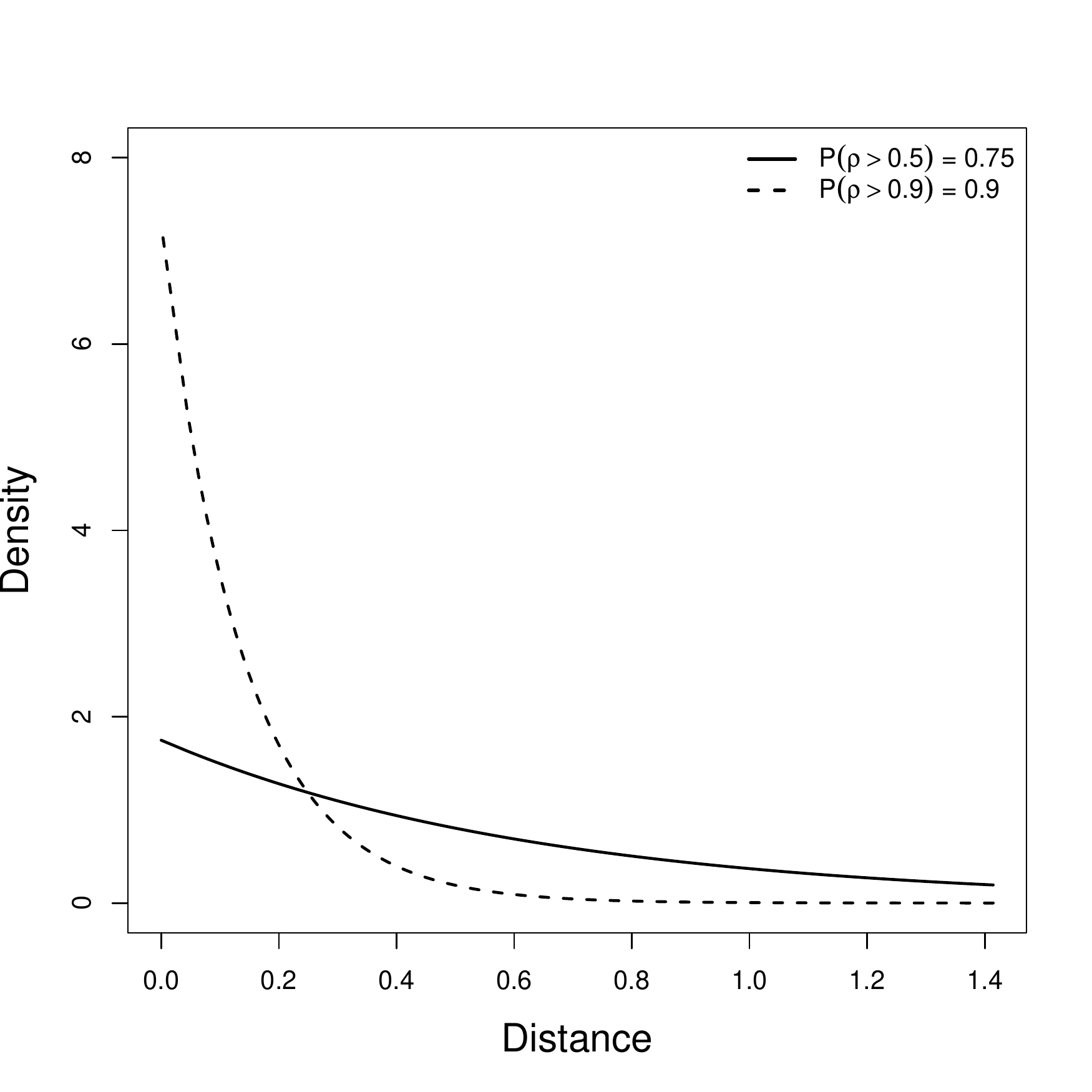}
}
\caption{Left panel: PC prior for $\rho$ under the AR(1) model; the base model is $\rho=1$. 
Right panel: the same PC prior plotted in the distance scale, $d(\rho)$; the base model is at $d(\rho)=0$.}
\label{fig:pc-rho-ar1}
\end{figure}

The user can incorporate information on his/her prior belief about the size of the correlation parameter by setting $U$ and $a$ so that $\mathbb{P}(\rho>U)=a$. To work out $\theta$ the equation 
\[
\frac{1-\exp(-\theta\sqrt{1-U})}{1-\exp(-\sqrt{2}\theta)}=a, \quad a > \sqrt{(1-U)/2}
\]
 needs to be solved numerically for $\theta$ as in the unstructured case. The PC prior in (\ref{eq:pc-rho-ar1}) is illustrated in Figure~\ref{fig:pc-rho-ar1} left panel.

\begin{figure}
\centerline{
\includegraphics[width=.5\textwidth]{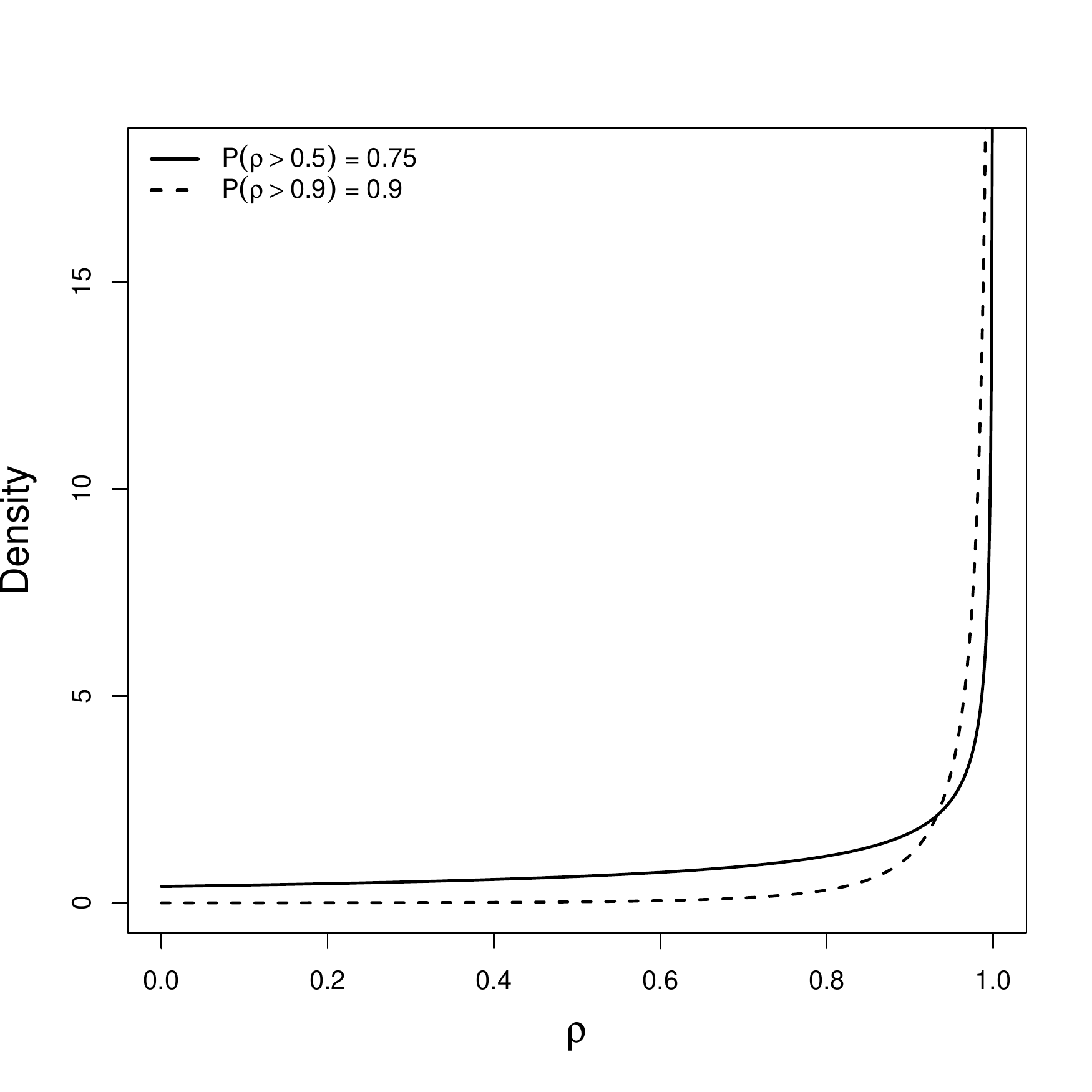}
\includegraphics[width=.5\textwidth]{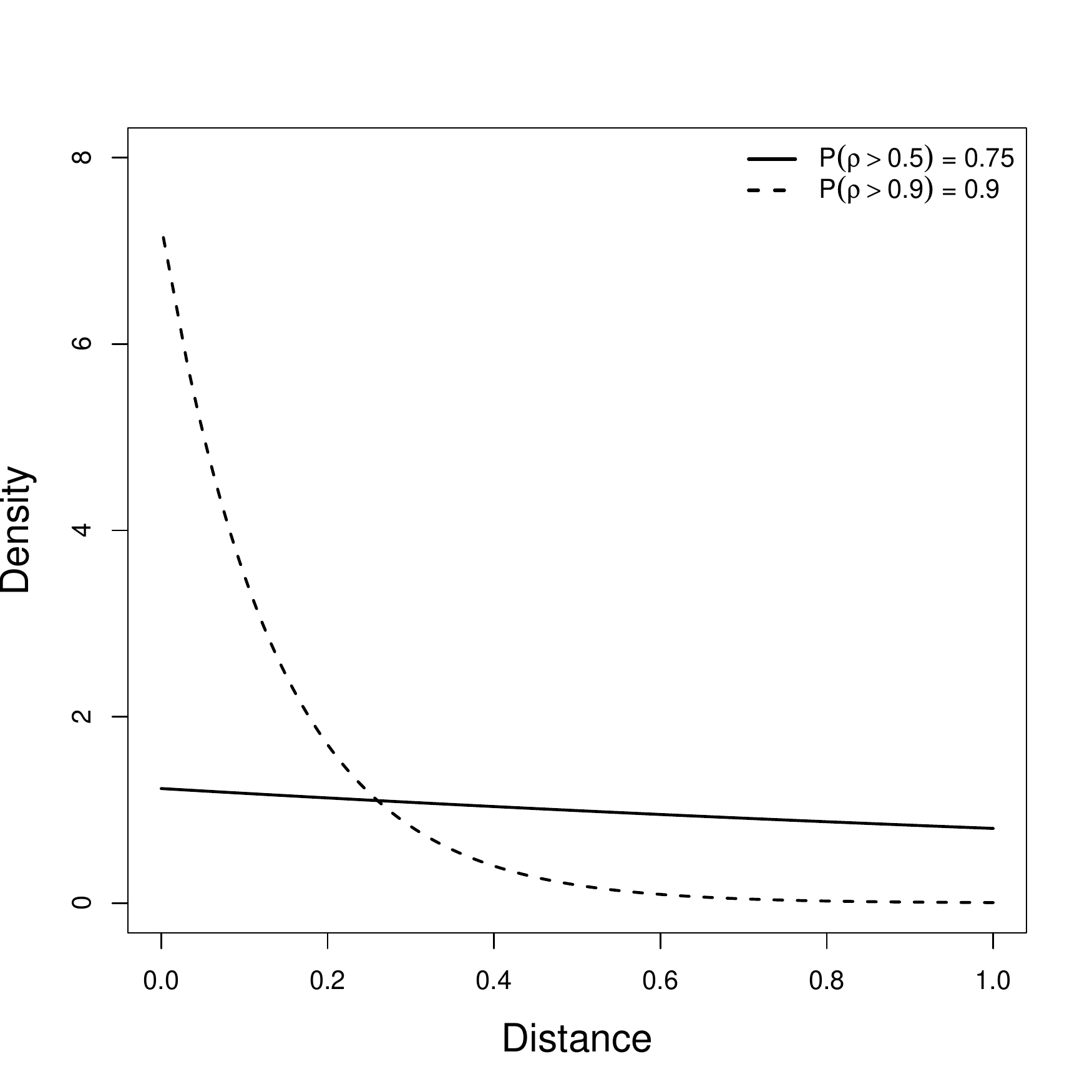}
}
\caption{Left panel: PC prior for $\rho$ under the exchangeable model; the base model is $\rho=1$.
Right panel: the same PC prior plotted in the distance scale, $d(\rho)$; the base model is at $d(\rho)=0$.}
\label{fig:pc-rho-exch}
\end{figure}

\subsubsection{Random walk model of order one and two}
\label{sec:pc:rw}
In the case of Model~(\ref{eq:igmrf}), the amount of deviation from the base model depends on $\tau$, with base model at $\tau = \infty$. \cite{pcprior} derive the PC prior for $\tau$ as a $\text{Gumbel}(1/2, \theta)$ type 2 distribution
\begin{equation}
\pi(\tau)= \frac{\theta}{2} \tau^{-{3}/{2}} \exp\left( -{\theta}/{\sqrt{\tau }}\right), \quad \tau>0,\theta>0.
\label{gumbel}
\end{equation}

\begin{figure}
\centerline{
\includegraphics[width=.5\textwidth]{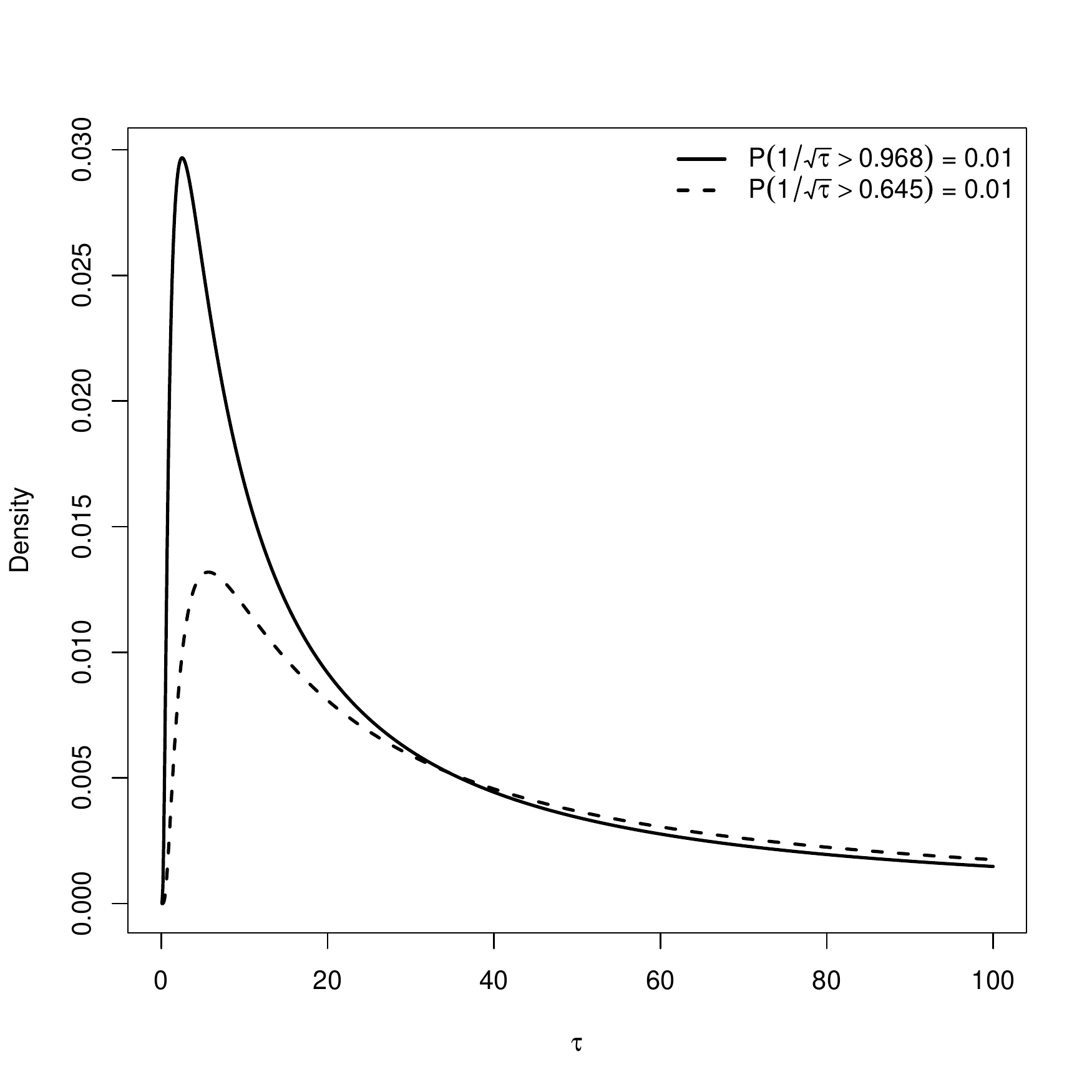}
\includegraphics[width=.5\textwidth]{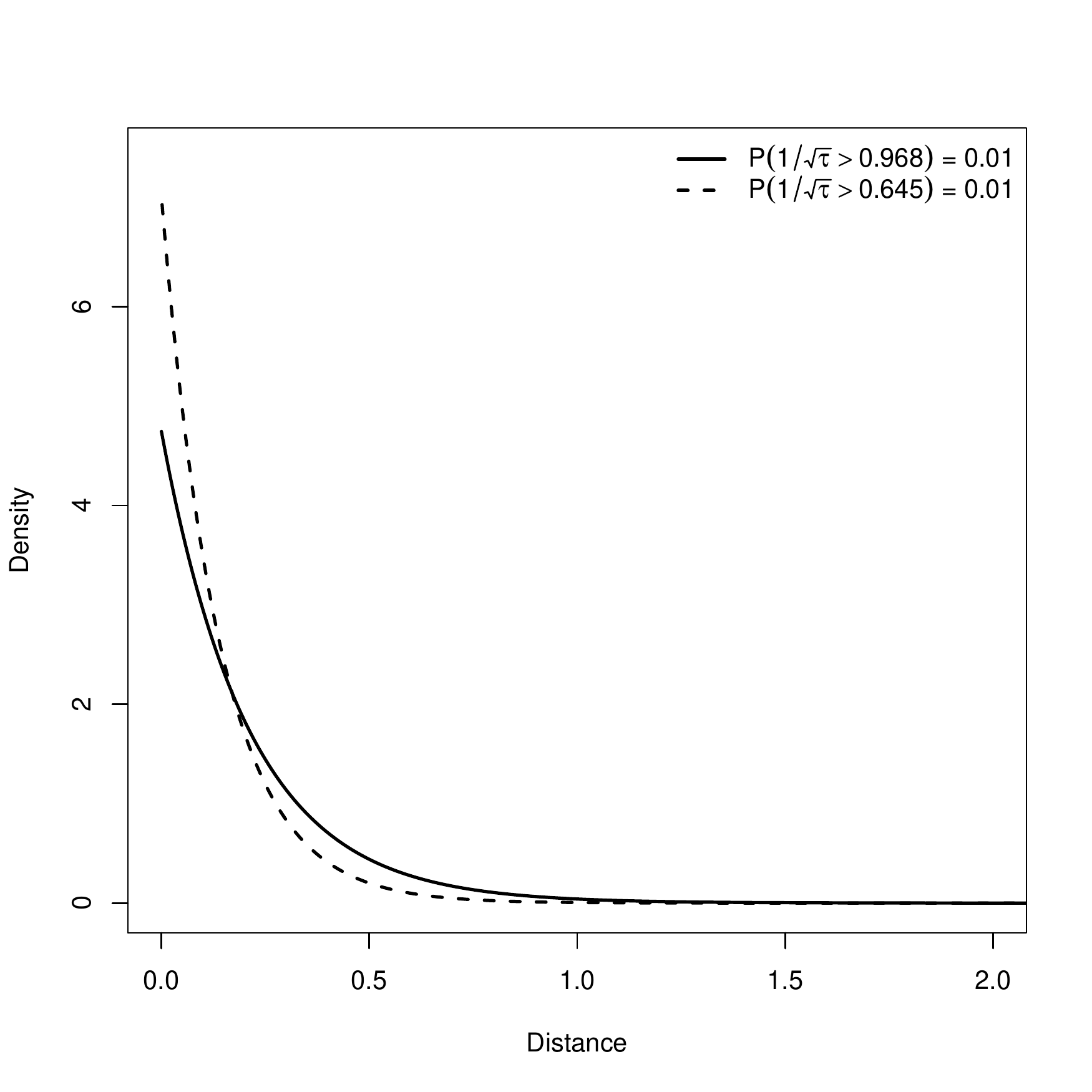}
}
\caption{Left panel: PC prior for $\tau$ under the RW model; the base model is $\tau=\infty$.
Right panel: the same PC prior plotted in the distance scale, $d(\tau)$; the base model is at $d(\tau)=0$.}
\label{fig:pc-tau-rw}
\end{figure}

To derive the scaling parameter $\theta$, \cite{pcprior} suggest to bound the marginal standard deviation, $1/\sqrt{\tau}$. This way it is sufficient to specify $(U,a)$ and solve $\mathbb{P}(1/\sqrt{\tau} > U) = a$ for $\theta$, which gives $\theta=-\log(a)/U$.
To aid the user in specifying parameters $(U,a)$, \cite{pcprior} provide a general rule of thumb: ``the marginal standard deviation of $\bm \beta$ with $\bm{K=I}$, after the type-2 Gumbel distribution for $\tau$ is integrated out, is about $0.31U$ when $\alpha=0.01$'';
 e.g. if we think a standard deviation of approximately 0.3 is a reasonable upper bound for the random effect (i.e. the varying coefficient), we need to set $U=0.3/0.31=0.968$. The PC prior in (\ref{gumbel}) is illustrated in Figure~\ref{fig:pc-tau-rw} left panel.

\subsection{The structured case: spatial variation}
\label{sec:pc:space}

\subsubsection{Areal spatial variation}
\label{sec:pc:areal}

It is clear from Eq.~(\ref{eq:icar}) that the ICAR model can be seen as a RW1 model (Eq.~(\ref{eq:igmrf}) with \texttt{rank}$(\bm K)=n-1$), and hence the PC prior for $\tau$ is (\ref{gumbel}) as in the previous section.

\subsubsection{Continuous spatial variation}
\label{sec:pc:geostats}

PC priors for the range and marginal variance parameters of a GRF with \emph{Mat\'{e}rn} covariance function have been derived by \cite{Fuglstad:2017}.
The joint PC prior for $(\tau,\phi)$ with base model at $\tau=\infty$, $\phi=\infty$:
\begin{equation}\label{eq:pc-matern}
	\pi(\tau,\phi)=\lambda_{\phi}\phi^{-2}\exp\left(-\lambda_{\phi}\phi^{-1}\right)\frac{\lambda_{\tau}}{2}\tau^{-3/2}\exp\left(-\frac{\lambda_{\tau}}{\sqrt{\tau}}\right), \quad \quad \tau>0,\phi>0
\end{equation}
where, once the user fixes $U_{\phi}$,$a_{\phi}$,$U_{\tau}$,$a_{\tau}$ such that $\mathbb{P}(\phi<U_{\phi})=a_{\phi}$, $\mathbb{P}(1/\sqrt{\tau}>U_{\tau})=a_{\tau}$ the parameters $\lambda_{\phi}$, $\lambda_{\tau}$ are calculated as
\[
	\lambda_{\phi}=-\log(a_{\phi})U_{\phi}, \quad \quad \lambda_{\tau}=-\frac{\log(a_{\tau})}{U_{\tau}}.
\]

\begin{figure}
\centerline{
\includegraphics[width=.5\textwidth]{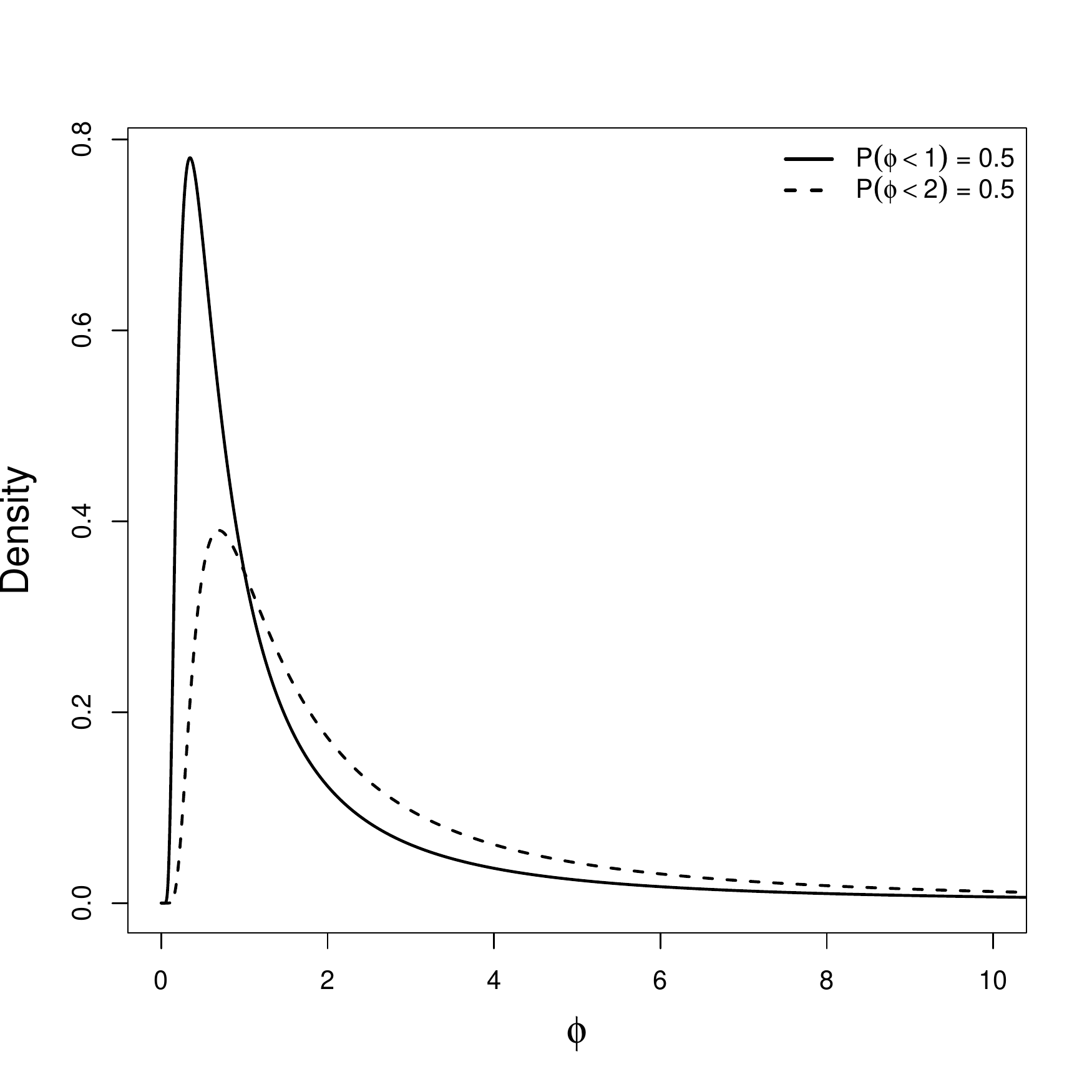}
\includegraphics[width=.5\textwidth]{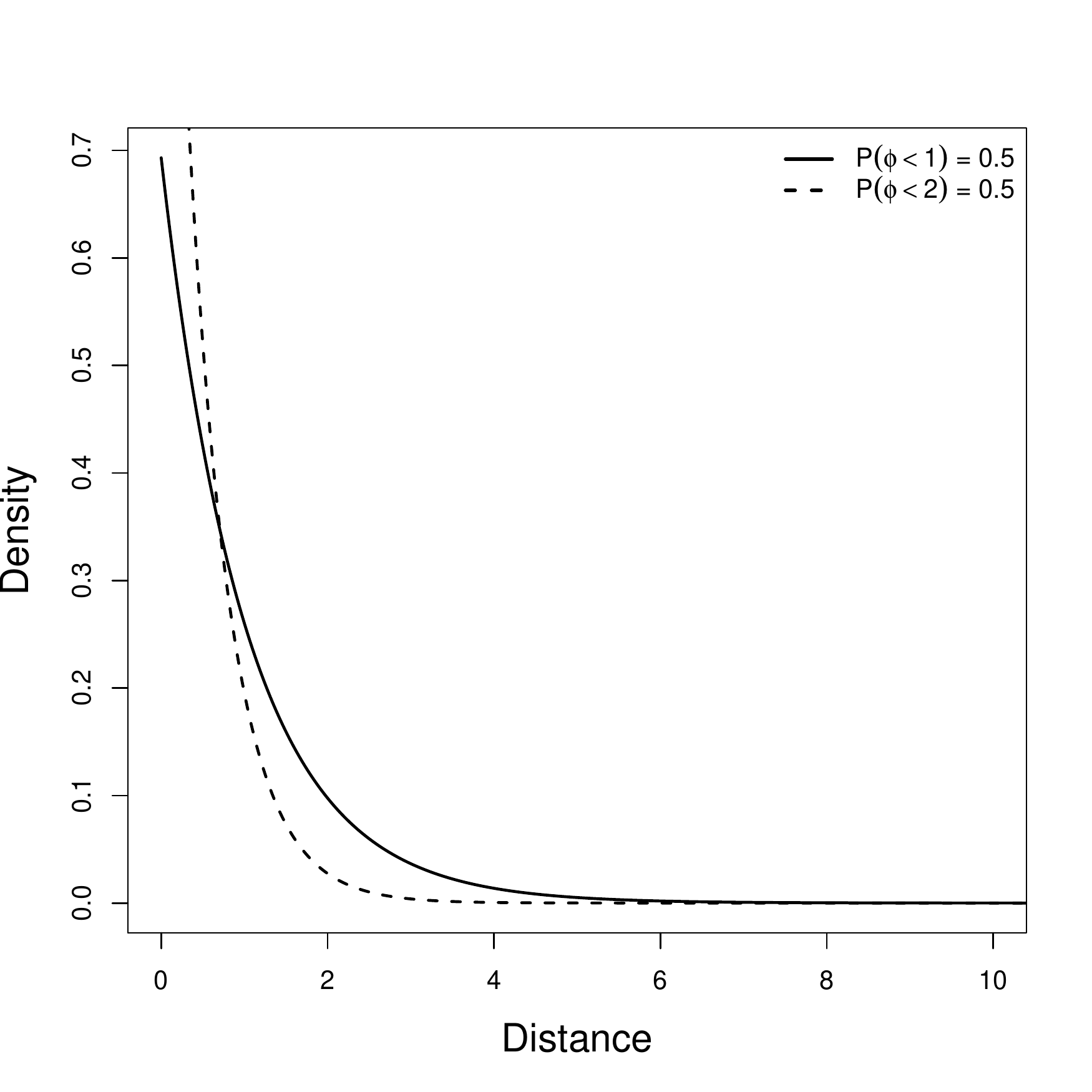}
}
\caption{Left panel: PC prior for range parameter $\phi$ of the \emph{Mat\'{e}rn} covariance function; the base model is $\phi=\infty$.
Right panel: the same PC prior plotted in the distance scale, $d(\phi)$; the base model is at $d(\phi)=0$.}
\label{fig:pc-range-matern}
\end{figure}

Note that, as the result of a convenient reparametrization (see appendix \ref{app:geostats}) the joint PC prior for $(\tau,\phi)$ factorizes as the product of the marginal densities. The PC prior for $\phi$ is illustrated in Figure~\ref{fig:pc-range-matern} left panel (while the PC prior for $\tau$ is the same as in Figure~\ref{fig:pc-tau-rw}).

\subsection{Properties of PC priors in the context of VCMs}
\label{sec:properties}

In Eq.~(\ref{eq:pc}) the PC prior is defined as an exponential distribution on the distance scale $d= \sqrt{2 \text{KLD}(f_1||f_0)}$, which implies two important properties.
First the exponential ensures constant rate penalization,
\begin{equation*}
\frac{\pi(d+\delta)}{\pi(d)} = \frac{\lambda \exp(-\lambda(d+\delta))}{\lambda\exp(-\lambda d)} = r^{\delta}, 
\end{equation*} 
where $r=\exp(-\lambda)$ is the constant decay rate. The relative change in the density for adding an extra $\delta$ does only depend on $\delta$, not on $d$. In many cases $d$ will not be an easy-to-interpret measure of complexity, thus the memory-less property becomes a practical device to penalize increasingly flexible models. (An example where the distance is well interpretable is for the case of independent Gaussian random effects, where $d$ corresponds to the marginal standard deviation of such random effects \citep{pcprior}). 

A second important property is that the mode of the PC prior is at distance $0$, meaning that PC priors naturally contract to the base model and prevent overfitting by construction. This is illustrated in the right panels of Figures~\ref{fig:pc-rho-exch},~\ref{fig:pc-rho-ar1},~\ref{fig:pc-tau-rw},~\ref{fig:pc-range-matern}, where each of the presented PC priors is displayed in their distance scale $d= \sqrt{2 \text{KLD}(f_1||f_0)}$.  \cite{pcprior} describe an \emph{overfitting prior} as a prior that places insufficient mass at $d=0$, suggesting that ``a prior overfits if its density in a sensible parametrization is zero at the base model'' (for justification of this choice see \cite{pcprior} section 2.4). The idea is the following: while priors that contract to $d=0$ avoid overfitting because they always give a chance for the base model to arise in the posterior, priors that go to $0$ at $d=0$ may incur in overfitting issues, because they may drag the posterior away from the base model, even when the latter is the true one. As an example, the conjugate Gamma for $\tau$ is an overfitting prior \citep{pcprior}. We believe this property is very important, as in the context of VCMs it is advisable to use priors that allow the \emph{constant coefficient} to arise in the posterior, unless data show evidence for a varying coefficient.

\subsubsection{Comparison with other priors}

\begin{figure}
\centerline{
\includegraphics[width=.5\textwidth]{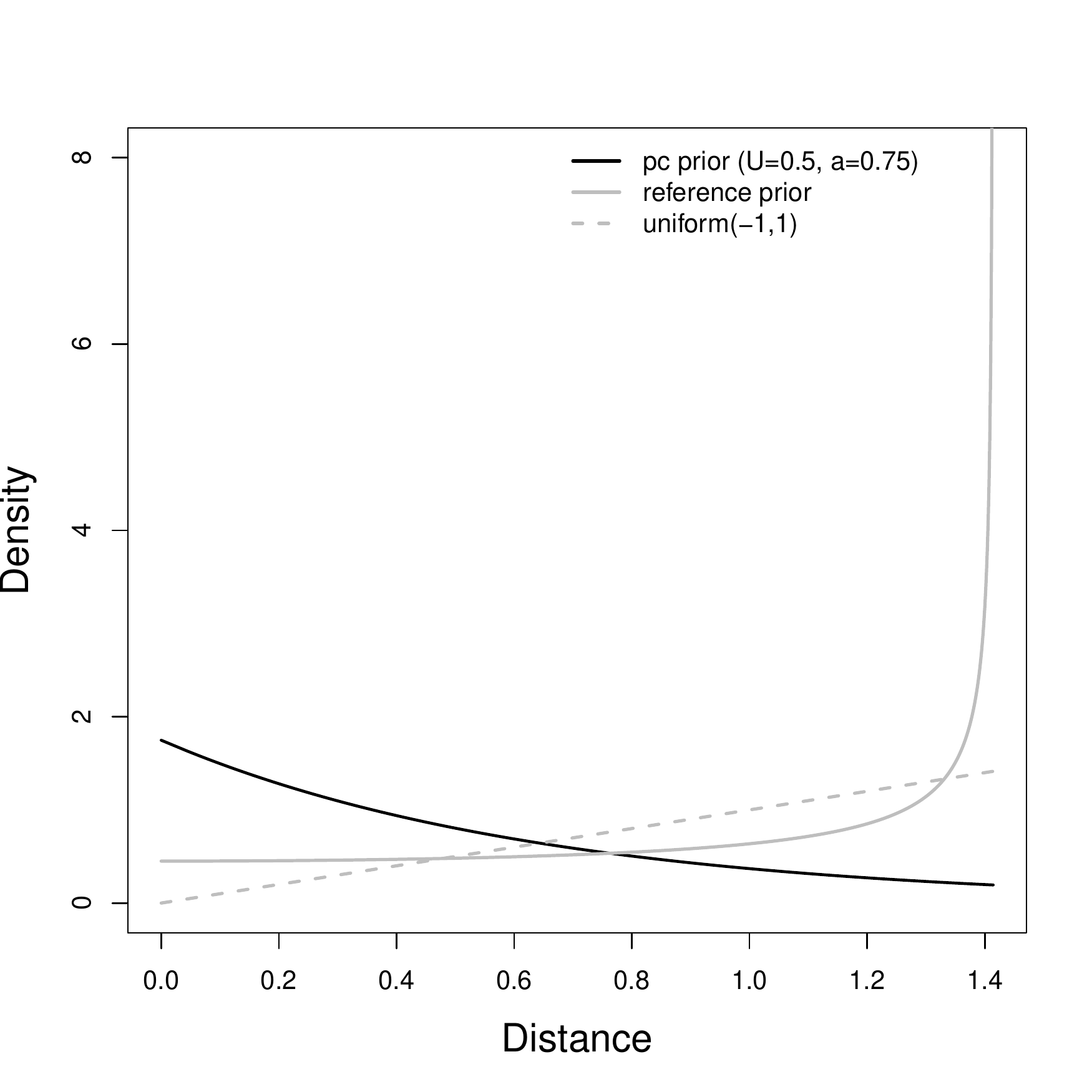}
}
\caption{The PC prior, reference and uniform prior for for the lag-one correlation $\rho$ of an AR1 model plotted in the distance scale.}
\label{fig:pc-compare-priors-ar1}
\end{figure}

Plotting priors on the distance scale is a useful tool to judge the behaviour near the base model. 
Figure~\ref{fig:pc-compare-priors-ar1} displays three different priors for the lag-one correlation $\rho$ of an AR1:  the PC prior in Eq. (\ref{eq:pc-rho-ar1}), the reference prior \citep{barnd, berger-yang} and the uniform on $(-1,1)$. All priors are plotted in the distance scale. 
The behaviour near the base model attained by the three priors is very different. The PC prior contracts to the base model as it peaks at minimum distance from the base model. The reference and uniform priors contract to the most complex model as they peak at maximum distance, $d=\sqrt{2}$. The uniform prior is the one that assigns less mass around $d=0$ and it overfits according to the informal definition in \cite{pcprior}. A simulation study was conducted to investigate performance of the priors depicted in Figure~\ref{fig:pc-compare-priors-ar1}. We considered a varying coefficient modelled as an AR1, focusing on two relevant cases: a first scenario (SC1) where the true VCM is close to the base model (i.e. a constant coefficient) and a flexible scenario (SC2) where the true VCM is far from the base model. See the supplementary material for more details on the simulation study and full discussion of results. In summary, it was found that the three priors perform equally well in SC2, while in SC1 the PC prior outperforms the other two, especially with noisy data. The uniform achieved poorest performance, which we presume is due to the fact that this prior forces overfitting in the sense of \cite{pcprior}. Our findings for the AR1 case are in line with several works comparing PC priors with other prior choices for the remaining models considered in this paper, see the supplementary material for details.

\section{Examples}
\label{sec:examples}

In the previous section we have shown how PC priors for varying coefficient models can be derived in a unified way regardless of the model assumed for the VC. Here we illustrate their application in two spatial examples where varying coefficient models are relevant.
All models are fitted within the \texttt{R-INLA} package \citep{martins-2013-new-inla} and the code is available in the supplementary material. The dataset used in example~\ref{sec:ex-baton} is freely available, while the data from the example in Section~\ref{sec:ex-sdo} cannot be published due to privacy issues, but the related \texttt{R-INLA} code is available using a simulated similar dataset. 

\subsection{PM$_{10}$ and hospital admissions in Torino, Italy}
\label{sec:ex-sdo}
The goal is to estimate the effect of PM$_{10}$ on the risk of hospitalization for respiratory causes using data on daily hospital admission from hospital discharge registers for the 315 municipalities in the province of Torino, Italy in 2004. In total, there are 12743 residents hospitalized for respiratory causes, aggregated by municipality and day. A reduced form of this dataset is available in the book by \cite{Blangiardo:2017}. 
Daily average temperature (Kelvin degrees) and particular matter PM$_{10}$ ($\mu g/m^3$) data are available at municipality level, the latter as estimates based on daily average PM$_{10}$ concentration \citep{Finazzi:2013}.

We consider the following model (all covariates are standardized), where the effect of PM$_{10}$ is allowed to vary spatially across municipalities:
\begin{eqnarray}
y_{i,t} & \sim & \text{Poisson}(E_{i,t}\exp(\eta_{i,t}))\nonumber \\
\eta_{i,t} & = & \alpha_t + u_i + \gamma \text{temp}_{i,t} + \beta_0 \text{PM}_{10,i,t} + \beta_i \text{PM}_{10,i,t} \label{eq:torino-vcm} \\
(\alpha_1, ..., \alpha_{366})^{\textsf{T}} & \sim & \text{cyclic RW2}(\tau_{\text{rw2}}) \label{eq:torino-rw2} \\
(u_1, ..., u_{315})^{\textsf{T}} & \sim & \text{BYM}(\tau_{\text{bym}}, \gamma_{\text{bym}})\label{eq:torino-bym}\\
(\beta_1,\ldots,\beta_n)^{\textsf{T}} & \sim & \text{ICAR}(\tau_{\text{icar}})\label{eq:torino-icar}
\end{eqnarray}
where $y_{i,t}$ and $E_{i,t}$ are the observed and expected number of hospitalizations in municipality $i=1,\ldots,315$ and day $t=1,\ldots,366$ respectively and $\exp(\eta_{i,t})$ is the relative risk of hospitalization in municipality $i$ and time $t$. 
Temperature (temp) is introduced as a fixed effect, as it is well known to be a confounder for the relationship between air pollution and health. PM$_{10,i,t}$ is taken as the sum of estimated daily average concentrations in the three days before $t$, in region $i$. 

With our model we are able to disentangle the mean effect of PM10 ($\beta_0$) expressing the overall change in the posterior relative risk for $1\mu g/m^3$ PM10 increase, from the varying cofficient $\beta_i$ expressing the municipality-specific deviation from $\beta_0$. We impose a sum to zero constraint on the $\beta_i$'s in Eq.~(\ref{eq:torino-icar}) to ensure identifiability of $\beta_0$, with $\beta_0 \sim N(0,1000)$. 

The random effects (\ref{eq:torino-rw2}) and (\ref{eq:torino-bym}) capture residual temporal and spatial structure, respectively. The temporal random effects are assigned a RW2 wrapped on a circle to ensure a cyclic trend over time; in practice, this is achieved by using a circulant precision matrix that constrains the first and last random effects to be the same, i.e. $\alpha_1=\alpha_{366}$ (see \cite{rue-2005}, section 2.6.1 for details). The spatial random effect $u_i$ is the sum of two random effects associated to municipality $i$, one spatially structured and one spatially unstructured, as defined by the popular BYM (Besag, York and Molli{\'e}) model \citep{besag-york-mollie}. We follow the BYM parametrization introduced by \cite{riebler-2016} and use the PC priors derived therein for the two hyperparameters of the BYM: a marginal precision $\tau_{\text{bym}}$, that allows shrinkage of the risk surface to a flat field, and a mixing parameter $\gamma_{\text{bym}} \in (0,1)$, that handles the contribution from the structured and unstructured components. For ease of notation, in (\ref{eq:torino-bym}) we skip all the details and refer the reader to \cite{riebler-2016}, formula (7). 

Table~\ref{table:tab-pcprior-parameter-sdo} summarizes the selected $U$ and $a$ for all PC priors. We can use the practical rule of thumb described at the end of Section~\ref{sec:pc:rw} to set an upper bound for the standard deviation. 
Weak prior knowledge suggests an upper bound for the marginal standard deviation approximately equal to $1,3$ and $0.1$ for the temporal trend ($\alpha_t$), the spatial component ($u_i$) and the VC ($\beta_i$), respectively. For instance, the choice of $U=0.1$ for $\beta_i$ is to be interpreted as: there is roughly $95\%$ probability that $\beta_i \in (e^{-0.1 \cdot 1.96},e^{0.1 \cdot 1.96})$, i.e. there is little chance that the deviation in increased relative risk (associated to $1\mu g/m^3$ increase in PM$_{10}$) is larger than $1.2$ in a given area. 

The change in the posterior relative risk for a $10\mu g/m^3$ increase in PM$_{10}$ is 1.002 (with 95\% credible interval (0.998,1.006)).
Figure~\ref{fig:ex2-fit} (panel a) displays the posterior mean for $\beta_i$, i.e. the municipality specific deviations (in the linear predictor scale) from the mean effect of $PM_{10}$. Panel (b) in Figure~\ref{fig:ex2-fit} shows the posterior probability of an increased risk associated to pollution, demonstrating that changes in the varying coefficients across municipalities may only be substantial in the municipality of Turin (the \emph{hotspot} in the south-east area). Looking at the prior vs posterior in Figure~\ref{fig:ex2-learning-sensitivity} (a), we see that there seems to be some information in the data regarding $\tau_{\text{icar}}$ as prior and posterior are clearly separated.

From an epidemiological point of view, there seems to be two possible explanations for a spatially-varying pollution effect. First, the result might be due to the effect of an unobserved confounding variable which is not captured by the random effects in the model. Second, the PM$_{10}$ chemical composition might change substantially over space, so that the PM$_{10}$ may be more or less dangerous for people, according to where they live.

\subsubsection*{Sensitivity analysis}
\begin{figure}
\centerline{
\includegraphics[width=.5\textwidth]{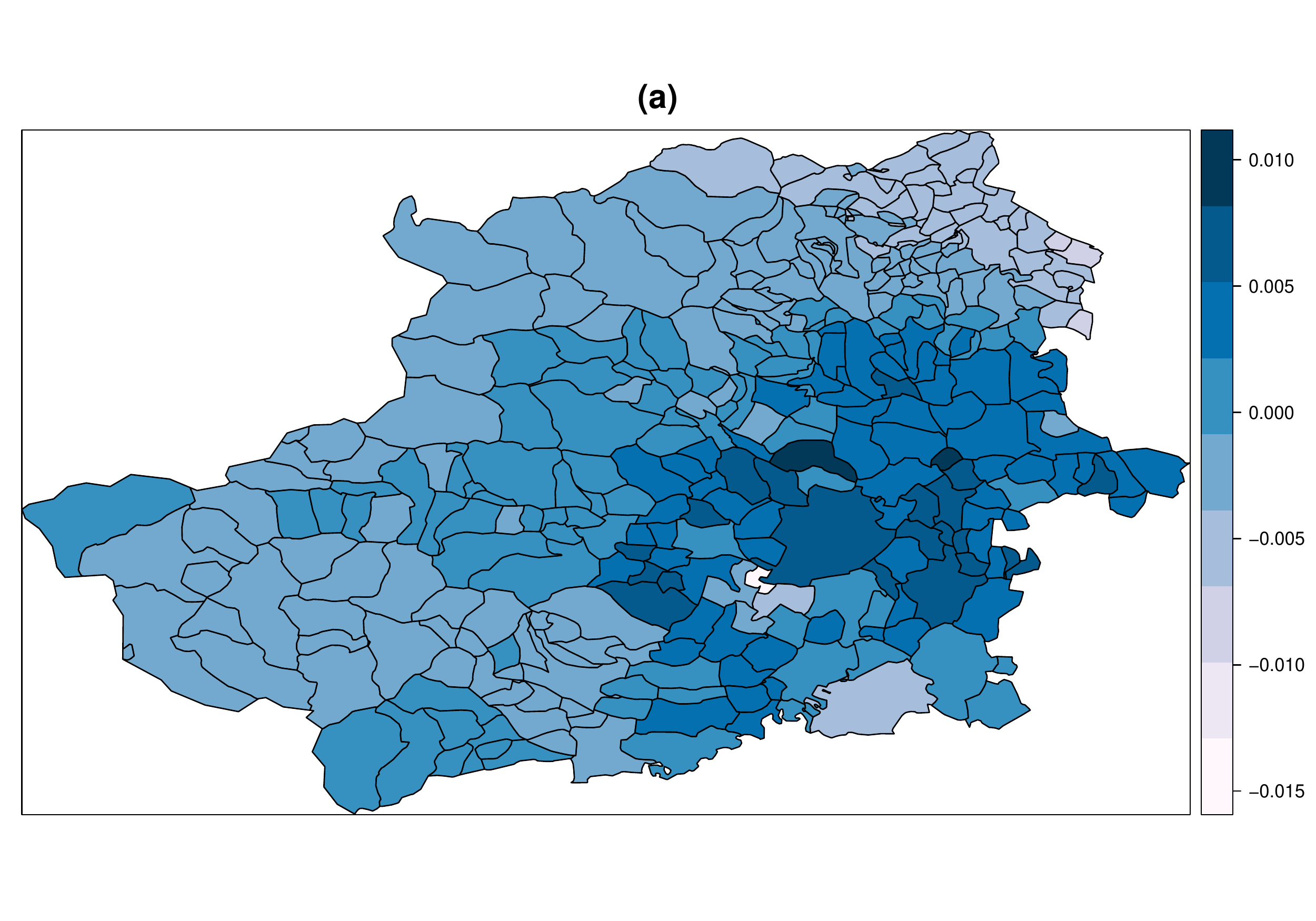}}
\centerline{
\includegraphics[width=.5\textwidth]{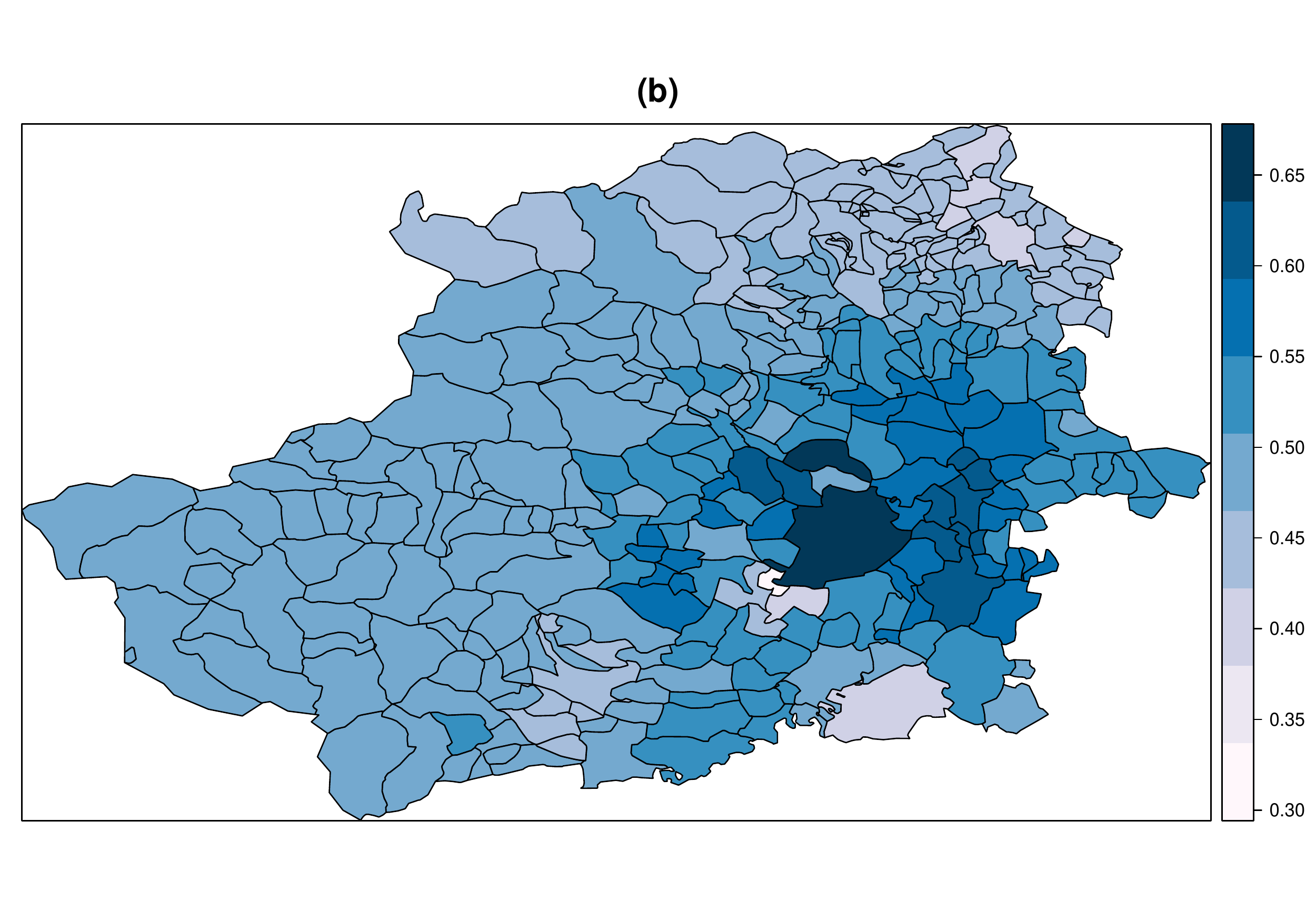}}
\caption{Posterior mean for the varying coefficients $\beta_i$ (panel a) and posterior probability $\mathbb{P}(\beta_i > 0|\bm y)$ (panel b).}
\label{fig:ex2-fit}
\centerline{
\includegraphics[width=.5\textwidth]{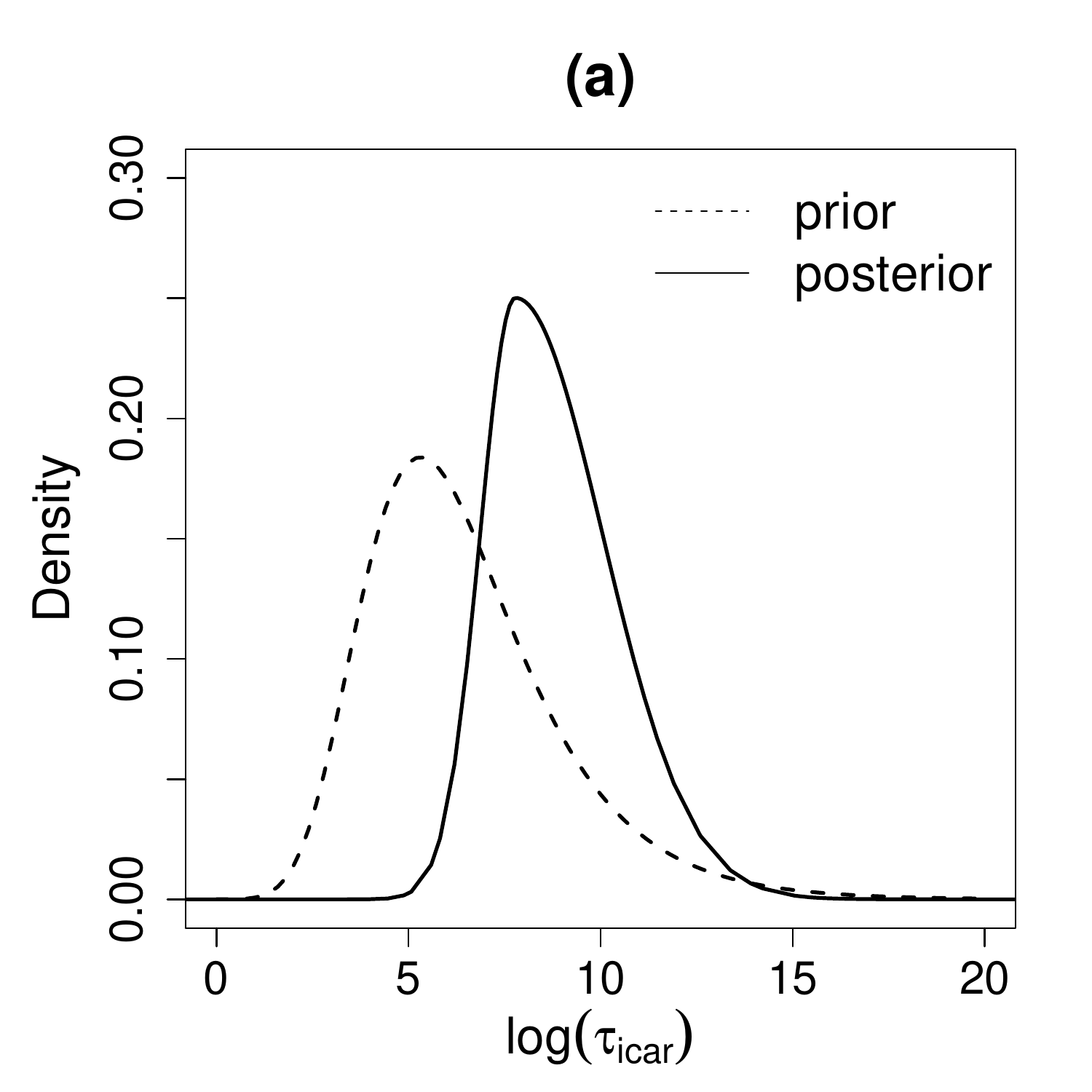}
\includegraphics[width=.5\textwidth]{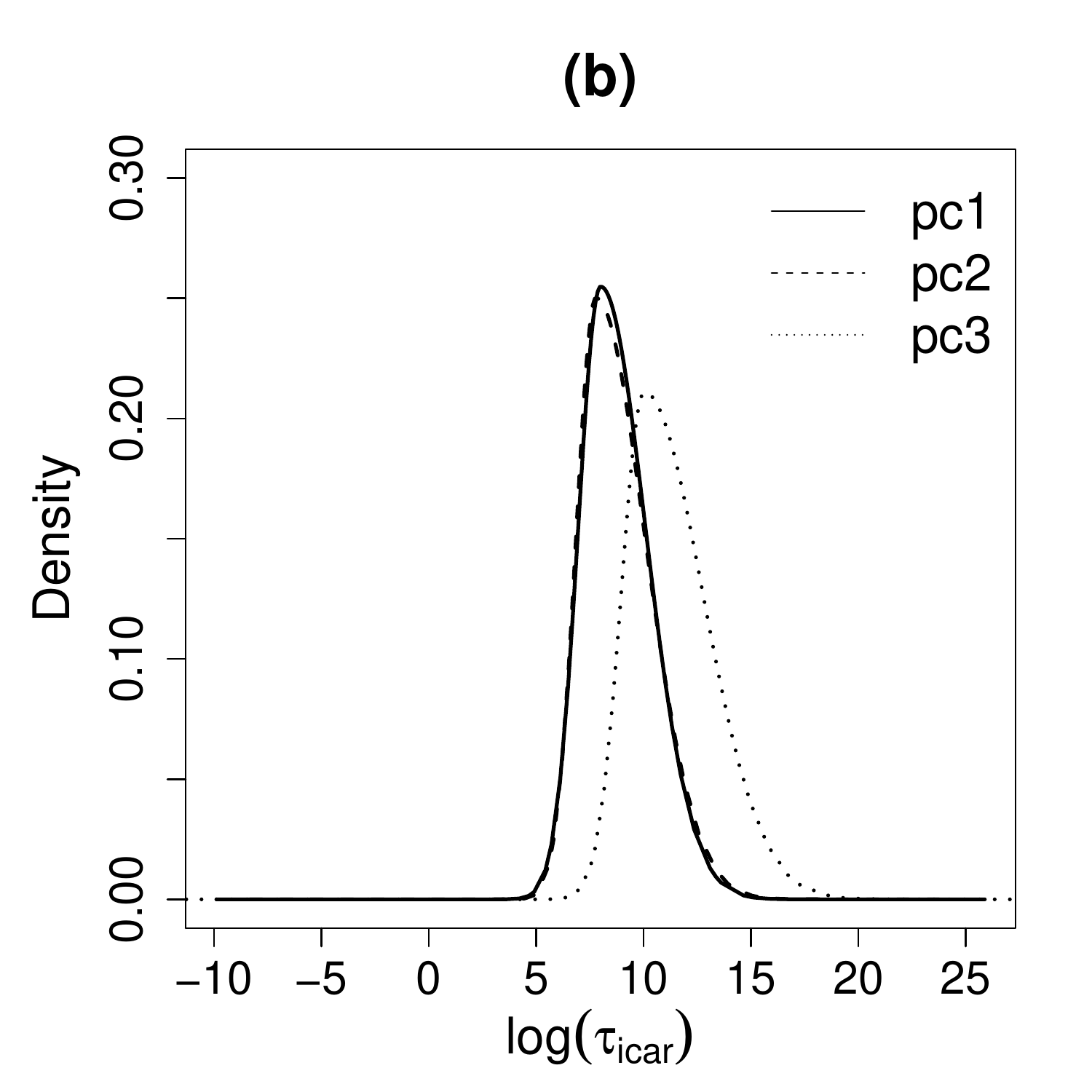}
}
\caption{Prior \emph{vs} posterior comparison for the precision parameter $\tau_\text{icar}$ as specified in Table~\ref{table:tab-pcprior-parameter-sdo} (panel a) and posterior for $\tau_{\text{icar}}$ for each setting in Table~\ref{table:tab-pcprior-sensitivity-sdo} (panel b).}
\label{fig:ex2-learning-sensitivity}
\end{figure}

An interesting question is how sensitive the model fit is to a change in the PC prior parameters $U,a$. Figure~\ref{fig:ex2-learning-sensitivity}(b) displays posterior distributions for $\tau_{\text{icar}}$ under three different settings (see Table~\ref{table:tab-pcprior-sensitivity-sdo}) with increasing penalty for deviating from the base model. There does not seem to be a great effect of $U$ on the posterior for $\tau_{\text{icar}}$ unless we impose a strong penalization for deviating from the base model (\texttt{pc3}).
In terms of posterior relative risks, results (not reported here) remain basically unchanged across the different prior scenarios, unless a prior for the precision that puts a lot of probability mass around the base model is used, in which case the risk pattern is more shrunk towards no variation. 

\begin{small}
\begin{table}[ht]
\centering
\caption{Summary of the PC prior parameters $U$ and $a$ used in model (\ref{eq:torino-vcm}) for the precisions ($\tau$) and the $\gamma$ parameter.
}\label{table:tab-pcprior-parameter-sdo}
\begin{tabular}{cccc}
  \hline
 PC prior & $\alpha_t $ (\text{rw2}) & $u_i $ (\text{BYM}) & $\beta_i $ (\text{ICAR})\\ 
  \hline
$\pi(\tau| U,a=0.01)$ & $U = 0.1/0.31$ & $U=3/0.31$ & $U=0.1/0.31$\\
$\pi(\gamma| U,a=0.5)$ & - & $U=0.5$ & -  \\
\hline
\end{tabular}
\end{table}
\end{small}

\begin{table}[ht]
\centering
\caption{Summary of the PC prior parameters $U$ and $a$ for $\tau_{\text{icar}}$ used in the sensitivity analysis for Model~(\ref{eq:torino-vcm}).}\label{table:tab-pcprior-sensitivity-sdo}
\begin{tabular}{cccc}
  \hline
 PC prior parameters & \texttt{pc1} & \texttt{pc2} & \texttt{pc3}\\ 
  \hline
$U$ & 1/0.31 & 0.1/0.31 & 0.01/0.31 \\
$a$ & 0.01 & 0.01 & 0.01  \\
\hline
\end{tabular}
\end{table}

A possible alternative could be to assume an exchangeable model for the varying coefficient. Given the large number of areas ($n=315$) we considered it was more natural to assume the varying coefficients to be spatially structured but for similar applications with a small number of areas an exchangeable model could be used.

\subsection{House prices in Baton Rouge, Louisiana}
\label{sec:ex-baton}
\begin{figure}
\centerline{
\includegraphics[width=.45\textwidth]{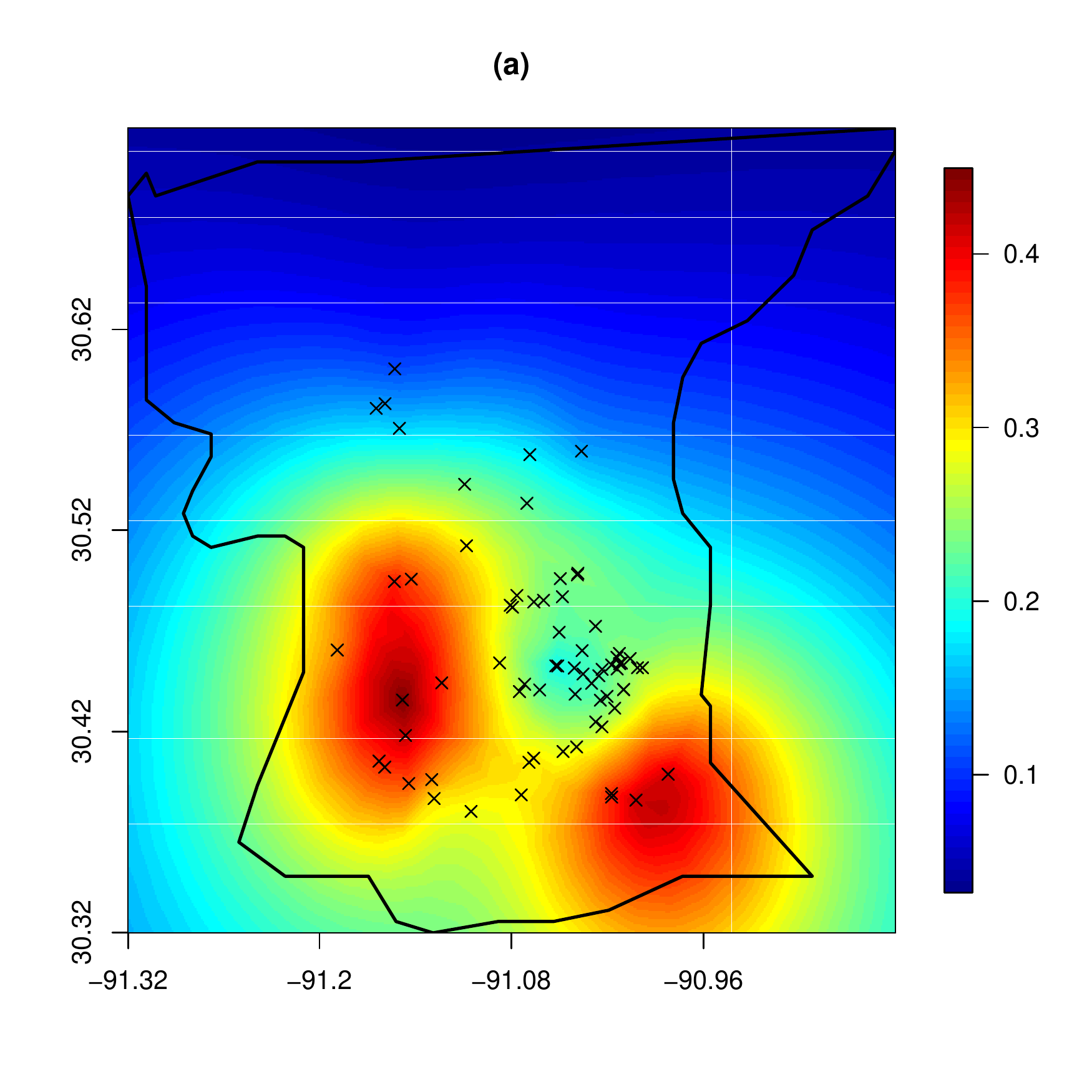}
\includegraphics[width=.45\textwidth]{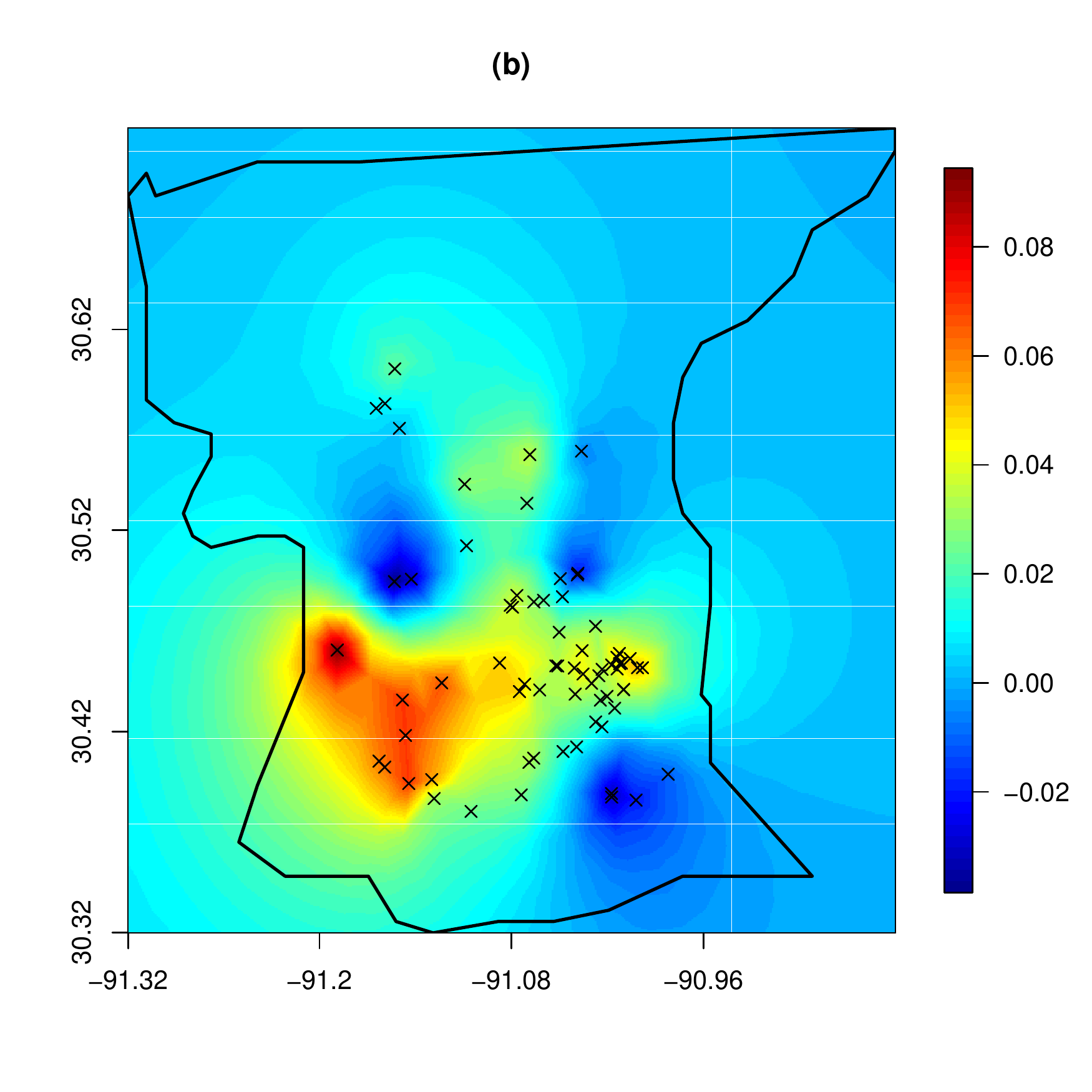}
}
\caption{Posterior mean for the varying coefficient of area $\beta_{\text{a},i}$ (panel a) and other area $\beta_{\text{b},i}$ (panel b). Observed locations are marked with a cross.}
\label{fig:ex3-VC-area-and-other-area}
\end{figure}

The dataset considered in this example is available in \cite{Banerjee:2015} and consists of selling prices (\$) of 70 single family homes in East Baton Rouge Parish, Louisiana, sold in June 1989. Living area (square feet) and other area (square feet) such as garden, garage, etc., are available as covariates, as well as the longitude (lon) and latitude (lat) coordinates. An extended version of this dataset is analyzed in \cite{Gelfand:2003}. The spatial locations of the houses sold can be seen in Figure~\ref{fig:ex3-VC-area-and-other-area}, along with the border delimiting the parish of East Baton Rouge. Even though the expectation is that bigger houses with a bigger external area are more expensive than smaller ones, location plays an important role in determining the price of a house. Hence, we allow for a spatially varying effect of living area (area) and other area (Oarea) in the following model (where the covariates have been standardized):

\begin{eqnarray}
\log(\text{price})_i & = &\alpha + \gamma_{lon}\text{long}_i+ \gamma_{lat}\text{lat}_i+\beta_{a,i}\text{area} +\beta_{b,i}\text{Oarea} + \epsilon_i + e_i \label{eq:baton-vcm}\\
 & & (\beta_{a,1}, ..., \beta_{a,n})^{\textsf{T}} \sim \mathcal{N}\left(\bm 0,\tau_a^{-1} \bm R(\phi_a) \right)\\
  & & (\beta_{b,1}, ..., \beta_{b,n})^{\textsf{T}} \sim \mathcal{N}\left(\bm 0,\tau_b^{-1} \bm R(\phi_b) \right)\\
 & & \epsilon_i\sim \mathcal{N}(0,\tau_{\epsilon}^{-1}\bm R(\phi_{\epsilon}))\\
 & & e_i\sim \mathcal{N}(0,\tau_e^{-1})
\end{eqnarray}
with $\bm R(\phi)$ as in Eq.~(\ref{eq:R-matern}). PC priors for the parameters of the \emph{Mat\'{e}rn} covariance functions $\phi_a,\tau_a$, $\phi_b,\tau_b$ and $\phi_{\epsilon},\tau_{\epsilon}$ were scaled as follows. The maximum distance between observed locations is 5.12, so we set $U_{\phi}=2$ and $a_{\phi}=0.5$ so that $\mathbb{P}(\phi<2)=0.5$ for all $\phi_a$, $\phi_b$ and $\phi_{\epsilon}$.  
Regarding the marginal standard deviation, prior knowledge on the scale of the response and of the covariates can be used to select $U_{\tau}$ and $a_{\tau}$ in a reasonable way; we set $U_{\tau}=0.1/0.31$ and $a_{\tau}=0.01$ for $\tau_a$ and $\tau_b$ (i.e. $\mathbb{P}(1/\sqrt{\tau}>0.1/0.31)=0.01$) and $U_{\tau}=0.4/0.31$ and $a_{\tau}=0.01$ for $\tau_{\epsilon}$ (i.e. $\mathbb{P}(1/\sqrt{\tau}>0.4/0.31)=0.01$).

The posterior varying coefficient estimates for living area and other area are shown in Figure~\ref{fig:ex3-VC-area-and-other-area}.
The effect of living area on log selling price (panel a) changes depending on location and is greater than that of other area (panel b); in particular, there are two hot-spots where the effect appears to be greatest. The one on the left roughly corresponds to the area where Baton Rouge, capital of the state of Louisiana, is located. The bottom right corner corresponds to a district where household income is greater than that of the region as a whole.

The effect of other area on log selling price also varies spatially as it can be seen in Figure~\ref{fig:ex3-VC-area-and-other-area} (b). In particular, the red spot on the left hand side is roughly located on downtown Baton Rouge, the historic area of the city. On the other hand, it seems plausible that for houses located on the outskirts of the main cities in the region, the variable other area does not have such a strong impact on house price.

A small sensitivity analysis (see Table~\ref{table:tab-pcprior-sensitivity-baton}), was carried out in order to assess the impact of varying $U$ and $a$. The results (not shown here) seldom vary unless a PC prior for $\tau$ with nearly all the mass concentrated on the base model (\texttt{pc.b}) is used (as already observed in example \ref{sec:ex-sdo}). 
In practice, it is not possible to disentangle the effect of the range and marginal variance of a GRF. This results in sometimes different posterior means and distributions for the parameters under the remaining prior specifications in Table~\ref{table:tab-pcprior-sensitivity-baton} but with essentially no differences in the fitted surfaces with respect to those shown in Figure~\ref{fig:ex3-VC-area-and-other-area}. Given this difficulty in separating the effect of parameters $\phi$ and $\tau$ we opted to use an informative prior for the marginal variance, where $U$ and $a$ can be set in a more intuitive way, and a less informative prior for the range parameter.

\begin{small}
\begin{table}[ht]
\centering
\caption{Summary of the PC prior parameters $U$ and $a$ used in model (\ref{eq:baton-vcm}) for the precisions ($\tau$) and the $\phi$ parameters in the sensitivity analysis.
}\label{table:tab-pcprior-sensitivity-baton}
\begin{tabular}{ccccccccccc}
  \hline
  & & & \multicolumn{8}{c}{\texttt{scenario}}\\
  \hline
  & \multicolumn{2}{c}{$a_i$} & \multicolumn{2}{c}{\texttt{pc.a}} & \multicolumn{2}{c}{\texttt{pc.b}} & \multicolumn{2}{c}{\texttt{pc.c}} & \multicolumn{2}{c}{\texttt{pc.d}}\\
  \hline
  & $a_{\phi}$ & $a_{\tau}$ & $U_{\phi}$ & $U_{\tau}$ & $U_{\phi}$ & $U_{\tau}$ & $U_{\phi}$ & $U_{\tau}$ & $U_{\phi}$ & $U_{\tau}$\\
  \hline
$\beta_{a,i}$ & $0.5$ & $0.01$ & $2$ & $1/0.31$ & $2$ &$0.01/0.31$ &$0.5$ &$0.1/0.31$ &$5$ &$ 0.1/0.31$\\
$\beta_{b,i}$ & $0.5$ & $0.01$ & $2$ & $1/0.31$ & $2$ &$0.01/0.31$ &$0.5$ &$ 0.1/0.31$ &$5$ &$0.1/0.31$\\
$\epsilon_i$ & $0.5$ & $0.01$ & $2$ & $4/0.31$ & $2$ &$0.04/0.31$ &$0.5$ &$0.4/0.31$ &$5$ &$ 0.4/0.31$\\ 
  \hline
\end{tabular}
\end{table}
\end{small}

\section{Discussion}
\label{sec:discussion}

Most of the vast literature available on varying coefficient models considers case-specific prior distributions depending on the type of effect modifier and pays little attention to the risk of overfitting entailed by the increased flexibility that these models offer. In this paper we present varying coefficient models as a single class of models and use a unified approach for setting priors, regardless of the model assumed on the coefficients. 
The definition of the varying coefficient model as a flexible extension of a simpler model \emph{calls for} eliciting priors that allow the simpler base model to arise. PC priors guarantee this; since the mode is at the base model, overfitting, a common aspect in complex hierarchical models, is avoided by construction. PC priors follow specific principles that remain unchanged no matter the model choice for the varying coefficient $\pi(\bm \beta|\xi)$. This means we can address prior specification for any varying coefficient model in the same way using well defined principles. 

We have illustrated the use of PC priors for varying coefficients in two different applications.
Whether the covariate is standardized or not obviously makes an impact on the scale of the varying coefficient, thus the user should be careful in defining the value $U$ for the precision parameter $\tau$ and change it accordingly if the scale of the covariate is transformed. In our experience the choice of $U$ does not impact much the posterior for $\bm \beta$, unless almost all the probability mass is assigned deliberately to the base model, i.e. unless an unreasonable prior is used, meaning a prior that is against our prior knowledge on the behaviour of the VC.
Building a prior on the distance from a base model allows the level of informativeness of the prior to be set according to the actual amount of prior information. In the VCM case, for instance, the PC prior can be set as a weakly informative prior for the precision as we usually have a reasonable guess on the scale of the varying coefficient (depending on the link function of the model, the scale of the data and of the covariate). .

With the aim of covering the most popular varying coefficient models we assumed $\bm \beta$ as a Gaussian process with marginal precision $\tau$ and focused on several structures for the covariance matrix, reflecting different behaviours for the varying coefficient. 
The class of models presented here does not include the scale mixture of normals, $\beta_t|\tau_t \sim \mathcal{N}(0,\tau_t^{-1}), t=1, \ldots,n$, where the precision parameter varies over the range of the effect modifier, leading to a more complex and computationally involved VCM. These models are useful in specific situations like sparse regression \citep{carvalho:2010} and adaptive smoothing 
\citep{scheipl:2009}. In a VCM setting, this kind of models could be useful when the varying coefficient is thought to be a smooth function with non constant degree of smoothness. 

To conclude, choice of the prior $\pi(\xi)$ is difficult in practice, because there is typically no prior information on the hyperparameters in hierarchical models. Moreover, the empirical information available to estimate the posterior for $\xi$ is less compared to that available for the parameters in the linear predictor. This means that the the prior for $\xi$ is deemed to have a large impact on the model, especially in cases where data are poorly informative. 
 In our opinion, this represents a further good reason for using PC priors in varying coefficient models, as we can be more confident that no overfitting takes place when there is not enough information in the data. Even though we do not know much at prior about suitable values for $\xi$, we often know exactly what a hyperparameter \emph{does} in terms of shrinkage to a simpler model.

\section*{Acknowledgements}
Maria Franco-Villoria and Massimo Ventrucci are supported by the PRIN 2015 grant project n.20154X8K23 (EPHASTAT) founded by the Italian Ministry for Education, University and Research.


\newpage

\appendix

\section{Appendix: Derivation of the PC prior}
\label{app:pcprior}

\subsection{The unstructured case}\label{app:unstructured}
The varying coefficient model in the exchangeable case is
\[
\begin{array}{l}
\eta_t = \alpha +  \beta_t  x_{t} \ \ \ \ \ \ \ t=1,...,n,\\
\bm \beta \sim \mathcal{N}(0,\bm R(\rho)),
\end{array}
\]
with 
\[
\bm R(\rho) = \left[
\begin{array}{cccccc}
  1 & \rho & \ldots  &       & \rho   \\
  \rho & 1  & \rho & \ldots    & \rho   \\
   \cdot &    &   \cdot&   &  \cdot   \\
   \cdot &    &  &   \cdot&  \cdot \\
    \rho &  \rho    & \ldots &  \rho & 1 \\
\end{array}
\right]
\]
and base model $\rho=1$ (i.e. $\beta_t=\beta \ \ \forall t$). To evaluate the distance from the base model we need to use a limiting argument and consider a fixed value of $\rho=\rho_0$ close to 1 under the base model.
For zero-mean multivariate normal densities, the KLD simplifies to:
\[
\text{KLD}(f_1(\rho)||f_0) =  \frac{1}{2}\left(tr(\bm\Sigma_0^{-1}\bm\Sigma_1)-n-\log\left(\frac{|\bm\Sigma_1|}{|\bm\Sigma_0|}\right)\right)
\]
where $\bm\Sigma_0 = \bm R(\rho_0)$ and $\bm\Sigma_1 = \bm R(\rho)$, $\rho<\rho_0$, are the variance-covariance matrices under the base and flexible model respectively. In this case, the KLD:
\[
\text{KLD}(f_1(\rho)||f_0) = \frac{1}{2}\left(\frac{n(1+(n-2)\rho_0 - (n-1)\rho \rho_0)}{(1-\rho_0)((n-1)\rho_0+1)} - n - \log \frac{(1+(n-1)\rho)(1-\rho)^{n-1}}{(1+(n-1)\rho_0)(1-\rho_0)^{n-1}}\right)
\]
Considering the limiting value as $\rho_0 \to 1$, the distance
\[
	d(\rho)= \lim_{\rho_0 \to 1} \sqrt{2\text{KLD}(f_1(\rho)||f_0)}=\lim_{\rho_0 \to 1} \sqrt{\frac{(n-1)(1-\rho)}{1-\rho_0}}=c\sqrt{1-\rho}, \quad 0 \leq \rho < 1
\]
for a constant $c$ that does not depend on $\rho$. 
Since $0 \leq d(\rho) \leq c$, assigning a truncated exponential with rate $\lambda$ on $d(\rho)$ we have
\[
\pi(d(\rho)) = \frac{\lambda \exp(-\lambda c \sqrt{1-\rho})}{1-\exp(-\lambda c )}, \ \ \ \ \ \ \ \ \ \ 0 \leq d(\rho) \leq c, \quad \lambda>0. 
\]
Reparametrizing $\theta=\lambda c$ leads to the PC prior for $\rho$:
\[
	\pi(\rho)=\frac{\theta\exp(-\theta\sqrt{1-\rho})}{2\sqrt{1-\rho}(1-\exp (-\theta))}, \quad 0 \leq \rho < 1,\quad \theta>0.
\]

\subsection{The autoregressive model of first order}\label{app:AR1}
The varying coefficient model in the AR1 case is
\[
\begin{array}{l}
\eta_t = \alpha +  \beta_t  x_{t} \ \ \ \ \ \ \ t=1,...,n,\\
\bm \beta \sim \mathcal{N}(0,\bm R(\rho)),
\end{array}
\]
with $\bm R(\rho)_{ij}=(\rho^{|i-j|})$ and base model $\rho=1$. 
Using a limiting argument similar to that of Appendix~\ref{app:unstructured}, the distance to the base model is
\begin{equation}
d(\rho) = c \sqrt{1-\rho}, \ \ \ \ \ \ \ \ \ \ |\rho| < 1
\label{eq:d-nochangeintime}
\end{equation}
where $c$ is a constant. Note that (\ref{eq:d-nochangeintime}) is upper bounded, $0 \leq d(\rho) \leq c\sqrt{2}$, so that the PC prior for $d(\rho)$ is
\[
\pi(d(\rho)) = \frac{\lambda \exp(-\lambda c \sqrt{1-\rho})}{1-\exp(-\lambda c \sqrt{2})}, \ \ \ \ \ \ \ \ \ \ 0 \leq d(\rho) \leq c \sqrt{2}, \quad \lambda>0.
\]
Reparametrizing $\lambda=\theta/c$ and using the change of variable formula it follows that the PC prior on the $\rho$ scale is \citep{sorbye-2016} 
\[
\pi(\rho) = \frac{\theta \exp(-\theta \sqrt{1-\rho})}{2\sqrt{1-\rho}(1-\exp(-\sqrt{2}\theta)) }, \ \ \ \ \ \ \ \ \ \ |\rho| <1, \quad \theta>0. 
\]

\subsection{Random walk model of order one and two}\label{app:rw}
The varying coefficient has a joint distribution given by
\[
	\bm \beta \sim \mathcal{N}(\bm 0, \tau^{-1} \bm K^{-1})
\]
with $\bm K$ symmetric semi-positive definite matrix. Let $f_0=\pi(\bm \beta| \tau_0=\infty)$ and $f_1=\pi(\bm \beta| \tau)$ denote the base and flexible models, with precisions $\tau_0$ and $\tau$, respectively. 
\cite{pcprior} show that $\text{KLD}(f_1 || f_0)$ goes to $\frac{\tau_0n}{2 \tau}$, for $\tau$ much lower than $\tau_0$ and $\tau_0 \to \infty$, so that $d(\tau)=\sqrt{2 \text{KLD}(f_1 || f_0)}= \sqrt{{\tau_0n}/{ \tau}}$ and $d(\tau)\sim \exp(\lambda)$, $\lambda>0$.  

By a change of variable and setting the rate $\lambda=\theta/\sqrt{n \tau_0}$, \cite{pcprior} derive the PC prior for $\tau$ as 
\begin{equation}
\pi(\tau)= \frac{\theta}{2} \tau^{-{3}/{2}} \exp\left( -{\theta}/{\sqrt{\tau }}\right), \quad \tau>0,\theta>0,
\end{equation}
which is a $\text{Gumbel}(1/2, \theta)$ type 2 distribution. 

\subsection{Continuous spatial variation}\label{app:geostats}
The spatially varying coefficient can be seen as a realization of a Gaussian random field (GRF)
\[
\bm \beta \sim \mathcal{N}\left(\bm 0,\tau^{-1}\bm R(\phi)\right)
\]
with \emph{Mat\'{e}rn} correlation function as in (\ref{eq:pc-matern}).
PC priors for the range and marginal variance parameters of a GRF with \emph{Mat\'{e}rn} covariance function have been derived by \cite{Fuglstad:2017}. Here we only summarize the main results on the computation of the PC prior, while for further details the reader is referred to \cite{Fuglstad:2017}.
Deriving PC priors for these parameters is more complex that in the previous situations considered in this paper due to the infinite-dimensional nature of GRFs. Following \cite{Fuglstad:2017} and setting $d=2$, parameters $\phi$ and $\tau$ are conveniently reparametrized as:
\[
	\kappa=\frac{\sqrt{8\nu}}{\phi} \quad \quad \psi=\sqrt{\tau^{-1}}\phi^{\nu}\sqrt{\frac{\Gamma(\nu + 1)4\pi}{\Gamma(\nu)}}
\]

Since the parameter $\psi$ depends on $\kappa$, the joint PC prior is built as $\pi(\psi,\kappa)=\pi(\kappa)\pi(\psi|\kappa)$, which can then be transformed into a joint PC prior for $(\phi,\tau)$. In this case, the base model corresponds to $\phi=\infty$ (or equivalently, $\kappa=0$), i.e. the spatial correlation is so strong that we have a constant field and $\tau=\infty$ ($\psi=0$), i.e. no marginal variance. 
The PC prior $\pi(\psi|\kappa)$ is built based on the observations available at $n$ locations, while the PC prior $\pi(\kappa)$ is based on the infinite-dimensional GRF to avoid a model-dependent prior; see \cite{Fuglstad:2017} for details.

The PC prior for $\kappa$:
\begin{equation}
	\pi(\kappa)=\lambda_{1}\exp\left(-\lambda_{1}\kappa\right),\quad \kappa>0,
\end{equation}
and $\lambda_{1} >0$. 
The user can set $U_1$ and $a_{1}$ such that $\mathbb{P}(\phi<U_1)=a_1$, so that $\lambda_{1}=-\left(\frac{U_1}{\sqrt{8\nu}}\right)\log(a_{1})$.

The PC prior for $\psi|\kappa$ follows an exponential distribution: 
\begin{equation}
	\pi(\psi|\kappa)=\lambda_{2}\exp(-\lambda_{2}\psi), \quad \psi>0
\end{equation}
where, as before, $\lambda_{2}>0$ can be selected based on the user-selected values $U_2$ and $a_{2}$ such that $\mathbb{P}(1/\sqrt{\tau}>U_2|\kappa)=a_2$, which leads to $\lambda_{2}(\kappa)=-\kappa^{-\nu}\sqrt{\frac{\Gamma(\nu)}{\Gamma(\nu+1)4\pi}}\frac{\log(a_{2})}{U_2}$.

The joint PC prior $\pi(\kappa,\psi)=\pi(\kappa)\pi(\psi|\kappa)$, and by a change of variable (setting $\lambda_{\phi}=\sqrt{8\nu}\lambda_1$ and $\lambda_{\tau}=\kappa^{\nu}\sqrt{\frac{\Gamma(\nu+1)4\pi}{\Gamma(\nu)}}\lambda_2$) it follows that the PC prior for $\tau,\phi$:
\begin{equation}
	\pi(\tau,\phi)=\pi(\phi)\pi(\tau|\phi)=\lambda_{\phi}\phi^{-2}\exp\left(-\lambda_{\phi}\phi^{-1}\right)\frac{\lambda_{\tau}}{2}\tau^{-3/2}\exp\left(-\frac{\lambda_{\tau}}{\sqrt{\tau}}\right) \quad \quad, \tau>0,\phi>0
\end{equation}
where, once the user fixes $U_{\phi}$,$a_{\phi}$,$U_{\tau}$,$a_{\tau}$ such that $\mathbb{P}(\phi<U_{\phi})=a_{\phi}$, $\mathbb{P}(1/\sqrt{\tau}>U_{\tau})=a_{\tau}$ the parameters $\lambda_{\phi}$, $\lambda_{\tau}$ are calculated as
\[
	\lambda_{\phi}=-\log(a_{\phi})U_{\phi}, \quad \quad \lambda_{\tau}=-\frac{\log(a_{\tau})}{U_{\tau}}.
\]

\end{document}